\documentclass[fleqn,10pt]{wlscirep}
\usepackage[utf8]{inputenc}
\usepackage[T1]{fontenc}

\usepackage{booktabs}
\usepackage{multirow}
\usepackage{pdflscape}
\usepackage{geometry}
\usepackage{graphicx}
\usepackage{makecell}
\usepackage{rotating} 
\usepackage{subcaption}
\usepackage{adjustbox}

\title{Beyond Distance: Mobility Neural Embeddings Reveal Visible and Invisible Barriers in Urban Space}

\author[1,4*]{Guangyuan Weng}
\author[5]{Minsuk Kim}
\author[5]{Yong-Yeol Ahn}
\author[1,2,3*]{Esteban Moro}
\affil[1]{Network Science Institute, Northeastern University, Boston, MA 02115, USA}
\affil[2]{Department of Physics, Northeastern University, 02115 Boston, USA}
\affil[3]{Media Lab, Massachusetts Institute of Technology, Cambridge, MA 02139}
\affil[4]{Khoury College of Computer Sciences, Northeastern University, Boston, MA 02115, USA}
\affil[5]{Luddy School of Informatics, Computing, and Engineering, Indiana University, Bloomington, IN 47405, USA}
\affil[*]{weng.g@northeastern.edu, e.moroegido@northeastern.edu}


\begin{abstract}
Human mobility in cities is shaped not only by visible structures such as highways, rivers, and parks but also by invisible barriers rooted in socioeconomic segregation, uneven access to amenities, and administrative divisions. Yet identifying and quantifying these barriers at scale and their relative importance on people's movements remains a major challenge. Neural embedding models, originally developed for language, offer a powerful way to capture the complexity of human mobility from large-scale data. Here, we apply this approach to 25.4 million observed trajectories across 11 major U.S. cities, learning mobility embeddings that reveal how people move through urban space. These mobility embeddings define a functional distance between places, one that reflects behavioral rather than physical proximity, and allow us to detect barriers between neighborhoods that are geographically close but behaviorally disconnected. We find that the strongest predictors of these barriers are differences in access to amenities, administrative borders, and residential segregation by income and race. These invisible borders are concentrated in urban cores and persist across cities, spatial scales, and time periods. Physical infrastructure, such as highways and parks, plays a secondary but still significant role, especially at short distances. We also find that individuals who cross barriers tend to do so outside of traditional commuting hours and are more likely to live in areas with greater racial diversity, and higher transit use or income. Together, these findings reveal how spatial, social, and behavioral forces structure urban accessibility and provide a scalable framework to detect and monitor barriers in cities, with applications in planning, policy evaluation, and equity analysis.
\end{abstract}
\begin{document}

\flushbottom
\maketitle

\vspace{-0.5cm}

\thispagestyle{empty}

\section*{Introduction}
Cities are social and opportunity hubs that foster diversity, social interaction, innovation, and economic growth \cite{schlapfer2014scaling, pentland2015social, Mouratidis2017Built, bettencourt_origins_2013}. However, cities are also characterized by persistent inequality in accessibility and mobility for their residents ~\cite{jacobs_death_1993,florida2017new}. The urban landscape is scarred by large visible physical barriers like natural obstacles, transportation infrastructure, or amenities, which constrain mobility and social connections ~\cite{appleyard1980livable,jacobs_death_1993}. At the same time, invisible demographic boundaries define by class, race, income, or lifestyles, create social barriers between groups that restrict everyday mobility between neighborhoods and groups, perpetuating segregation and isolation of individuals and neighborhoods \cite{wang2018urban,levy_triple_2020,xu_using_2025}. 

Despite their impact on urban mobility, there is no comprehensive study of visible and invisible barriers in urban environments due to their complex, heterogeneous, and uneven effect across socio-demographic and geographical context. Borders may selectively impact specific groups, neighborhoods or income levels and potential benefit others ~\cite{florida2017new, moro2021mobility, fraser2024great}. Physical barriers, like transport corridors or high-traffic roads, create significant mobility constrains, particularly affecting pedestrians and individuals with mobility challenges and short spatial scales\cite{jacobs_death_1993, jesus2022barrier, Matos2023The}. However, they may enable connectivity between areas at larger distances \cite{aiello_urban_2024}. The intricate interactions between barriers often amplify their collective impact on urban mobility \cite{Wessel2022Business, Kwate2013Retail}. For example, some highway barriers in the U.S. were deliberately design to reinforce racial segregation, linking their visible and invisible role in constraining mobility \cite{trounstine2018segregation}. In parallel, sociodemographic factors such as access to transit usage, race, income strongly affect mobility patterns, restricting visitation patterns, and accessibility to amenities \cite{wang2018urban,luo2016explore,Barbosa2020Uncovering,small_banks_2021}. Finally, some barriers are not easily recognized. While parks are traditionally
viewed as inclusive public spaces~\cite{cervero2017beyond}, evidence suggests they can function as socio-demographic barriers, limiting interactions across sociodemographic groups \cite{jacobs_death_1993,fraser2024great}. 
Traditional studies of mobility barriers have largely relied on small case-studies or geographies \cite{grannis_importance_1998,jacobs_death_1993,ananat_wrong_2011,shelton_social_2015,jesus2022barrier,anciaes2020comprehensive}. More recent work has begun to use large-scale mobility data to identify barriers at broader spatial scales~\cite{jin2021identifying, pinter2023quantifying, aiello_urban_2024,toth2021inequality}, but these studies have focused primarily on visible, physical obstacles. As a result, they often overlook invisible, demographic, or institutional factors. This focus limits our understanding of how different types of barriers interact to shape mobility and may prevent the identification of other important, previously unrecognized constrains. Without a more comprehensive perspective, conclusions about urban mobility and accessibility risk being incomplete or misleading.

\begin{figure*}[t!]
\centering
\includegraphics[width=\linewidth]{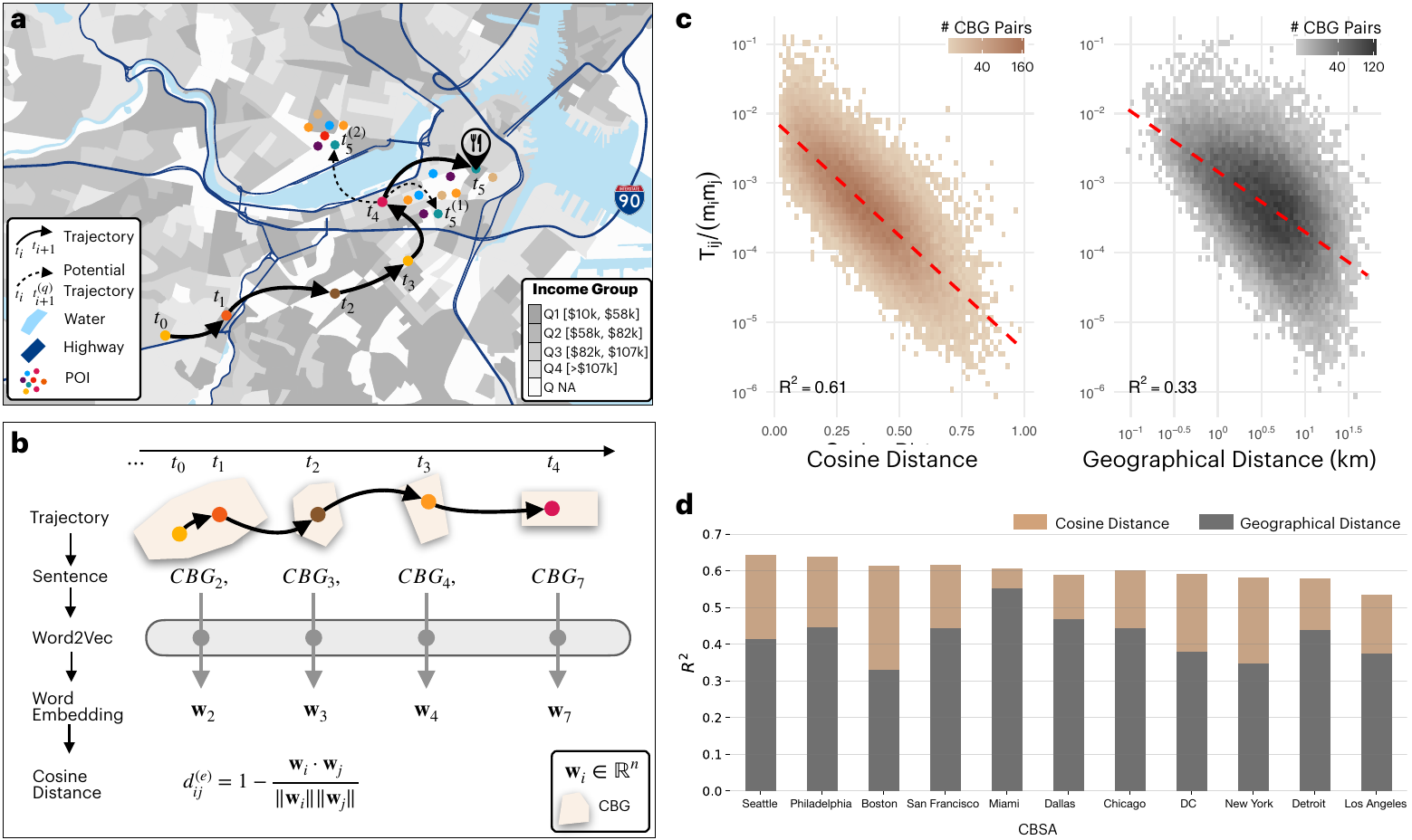}
\caption{
\textbf{Understanding mobility borders through human movement patterns and spatial embeddings.}
\textbf{(a)} 
Example trajectory in Downtown Boston showing how physical obstacles, income differences, and POI distribution influence destination choice. At time $t_5$, the individual selects a restaurant, bypassing closer options at $t^{(1)}_5$ and $t^{(2)}_5$ probably due to income differences and a large detour caused by the river. Maps created in R using TIGER shapefiles~\cite{tiger}.
\textbf{(b)} Schematic of the embedding process: CBGs are embedded using Word2Vec on individual trajectories, treating visited CBGs as words in a sentence. Resulting vectors $\mathbf{w}_i$ are used to compute embedding distances.
\textbf{(c)} Binned density plots for the Boston CBSA showing the relationship between normalized mobility flows between CBGs and embedding cosine distance (left) and geographic distance (right) in 2019. Red dashed lines show linear fits; embedding distance better explains flows ($R^2 = 0.61$ vs. $R^2=0.33$). 
\textbf{(d)} Bar plot of $R^2$ differences across 11 CBSAs in 2019, confirming the consistent advantage of embedding distances in modeling mobility flows. See SI Note~\ref{supp:cities_normalzied_flux} for other years.
}
\label{fig:1}
\end{figure*}

To address this gap by combining large-scale mobility data with neural embedding techniques that infer mobility barriers between urban areas from observed movement patterns. Fine-grained mobility data —from GPS traces to mobile phone records— has transformed the study of cities, enabling detailed analyses of mobility flows~\cite{gonzalez2008understanding,song2010modelling,pappalardo2015returners} revealing behavioral patterns in how individuals navigate urban environments~\cite{xu_using_2025,yang_identifying_2023,athey_estimating_2021,moro2021mobility,pappalardo_future_2023}. Building on these advances, we use large-scale anonymized, privacy-enhanced mobility trajectories across major U.S. cities to construct a representation of urban space that captures latent barriers to movement. Our embedding approach builds on recent methods used to model sequential human behavior~\cite{murray2023unsupervised,savcisens2024using,zhang_counterfactual_2024}, allowing us to identify barriers that are not evident from geography or physical characteristics alone. This framework moves beyond simple case studies or single-barrier analyses, enabling a systematic understanding of when, where, and for whom mobility barriers emerge in cities.

\section*{Results}
Our mobility data consists of anonymized, privacy-enhanced trajectories of 25.4 million devices across 11 metro areas \cite{cbsa} in the United States from 2019 to 2022 provided by Spectus. It is complemented by 2.5 million verified venues from Foursquare and Safegraph, as well as spatial data on major physical obstacles (highways, parks, water, etc.) from Open Street Map (see Methods and SI Note \ref{supp:data}). Fig.~\ref{fig:1}a illustrates an example trajectory in Downtown Boston, where an individual moves through several neighborhoods and multiple potential destinations. Although similar and equally distant amenities are available nearby, their final choice is shaped by a combination of visible barriers —such as rivers, highways, or amenities— and invisible ones, including income differences between neighborhoods or administrative boundaries. This example highlights the need to represent urban spaces and accessibility beyond geographical distance and visible structures, capturing the latent behavioral constraints that shape mobility, and assessing barriers at the points where movement decisions are made. 

To do this systematically across cities and at scale, we use a neural embedding-based approach that infers mobility barriers by measuring effective distances between locations from observed trajectories. To do that, we adopt the Word2Vec model due to its formal equivalence with the mobility gravity law under the skip-gram negative sampling formulation~\cite{murray2023unsupervised}. Mobility trajectories are treated as sequences of visited locations, allowing the model to learn vector embeddings $\mathbf{w}_i$ for each location $i$ based on co-occurrence patterns (See Fig.\ref{fig:1}b, Methods, and SI Notes~\ref{supp:traj}-\ref{supp:flow_pruning} for more details). For simplicity, we focus on trajectories aggregated at the Census Block Group (CBG) level, though our results are robust across other spatial resolutions (see SI Note~\ref{supp:h3_robustness}). Because Word2Vec is trained to predict location co-occurrence in mobility sequences, the resulting embeddings capture behavior-based distances $d_{ij}^{(e)}$ (i.e., cosine distance between $\mathbf{w}_i$ and $\mathbf{w}_j$) between CBGs that reflect actual movement patterns better than (Euclidean) geographic distances, $d_{ij}^{(p)}$. Importantly, this approach, contrary to other mobility models \cite{zipf1946p,pinter_neighborhoods_2023}, allow us to compute meaningful mobility distances even between CBGs with little or no direct flow, as long as they share similar mobility contexts. To evaluate the validity of these embeddings to describe real mobility, we model normalized flux using a gravity model $T_{ij}/m_i m_j \sim d_{ij}$ as a function of (Euclidean) physical or embedding distance. As shown in Fig.~\ref{fig:1}c–d, embedding distance has consistently higher explanatory power than geographic distance across cities and years, (e.g., $R^2 = 0.61$ vs. $R^2 = 0.33$ in Boston), validating our approach (see SI Note~\ref{supp:barrier_composition_by_year} and \ref{supp:cities_normalzied_flux} for full results across years and cities). We use Euclidean instead of travel distance in our comparison to make sure that physical barriers are reflected in the embedding, not excluded from the comparison.

\begin{figure*}[t]
\centering
\includegraphics[width=\linewidth]{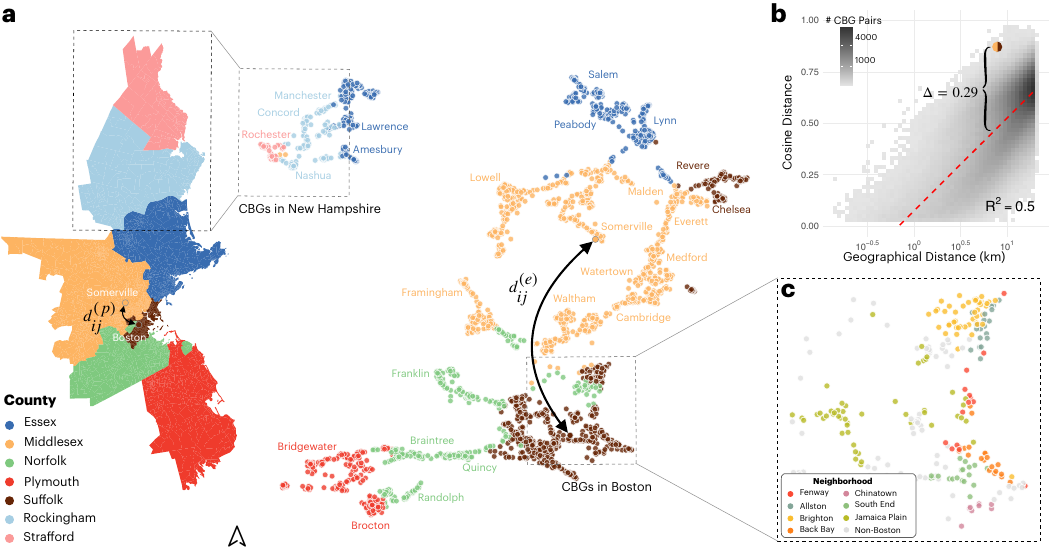}
\caption{
\textbf{Unsupervised embeddings uncover hidden mobility barriers and multi-scale urban structure.}
\textbf{(a)} Left: eographic map of the Boston metro area CBGs colored by county. Maps created in Python using TIGER shapefiles~\cite{tiger}. Center: UMAP projection of the embedding space, where each point is a CBG and colors indicate county membership. The projection shows different hierarchies of mobility at diverse spatial scales. The separation between New Hampshire and Massachusetts CBGs reflects administrative boundaries that constrain mobility. The annotated pair between Somerville and downtown Boston illustrates a mobility barrier: despite being geographically close, these locations are distant in embedding space.
\textbf{(b)} Binned density plot showing the relationship between geographic and embedding (cosine) distances. The red dashed line is the best linear fit ($R^2=0.50$). 
The barrier example from (a) is marked by a large positive residual, where observed mobility is much lower than predicted by the gravity model.
\textbf{(c)} Zoom-in of the UMAP mobility embedding on a cluster within Boston city. Colors indicate neighborhoods.
}
\label{fig:2}
\end{figure*}

The embeddings offer a data-driven representation of urban space that reflects how people move, rather than how places are situated geographically. To examine their structure, we use UMAP~\cite{McInnes2018} to reduce the high-dimensional embeddings to two dimensions. Because Word2Vec is trained to predict location co-occurrence in individual trajectories, embedding distances reflect mobility proximity: CBGs that are close in the embedding space and this UMAP projection tend to have frequent movement between them. As shown in Fig.~\ref{fig:2}a and Fig.~\ref{fig:2}c, the embedding space structure reveal a clear spatial hierarchy of the mobility within the Boston metropolitan area. At broader scales, the embedding still reproduces the global geographical structure of counties, states, and neighborhoods. However, at finer scales, it reveals patterns that diverge from geographic proximity. For example, CBGs in Strafford and Rockingham counties (NH) form distinct clusters, despite being adjacent to Essex County (MA), indicating limited cross-boundary mobility. Locally, areas like Somerville and Boston’s urban core appear far apart in the embedding space, despite geographic proximity. These divergences between geographic proximity and embedding distance highlight a key insight: proximity on a map does not necessarily imply connectivity between places. These mismatches between spatial closeness and mobility proximity reveal latent behavioral barriers.

\begin{figure*}[t]
\centering
\includegraphics[width=\linewidth]{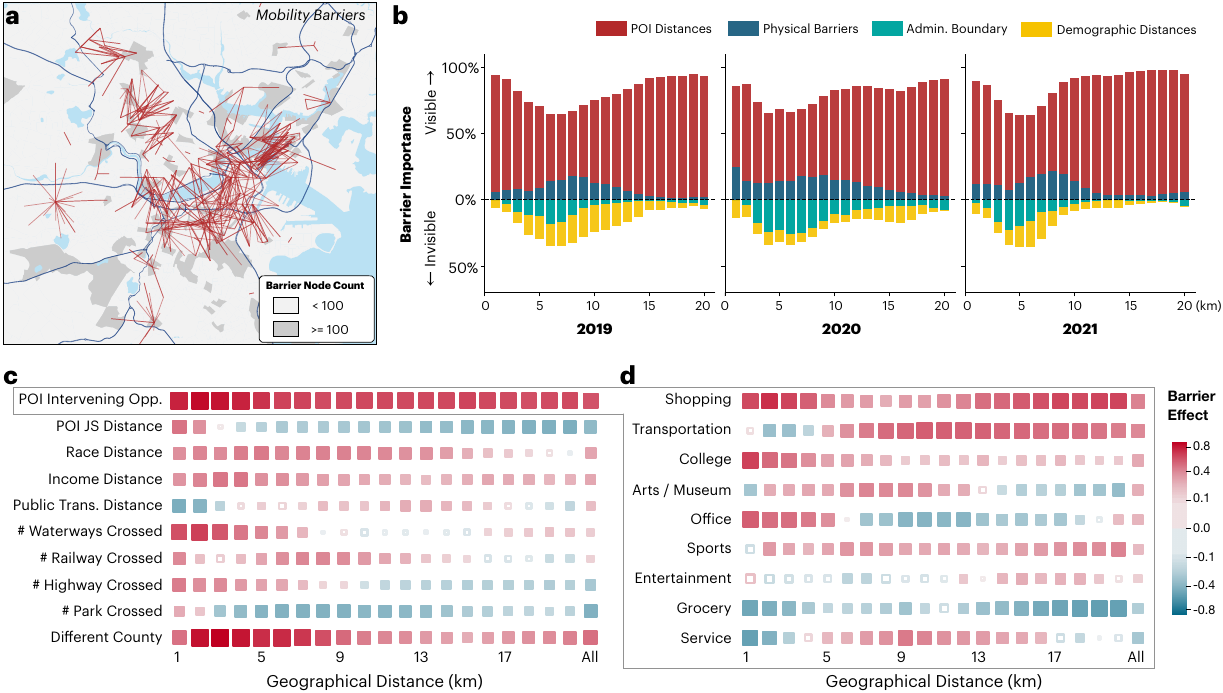}
\caption{
\textbf{Visible and Invisible Factors Shape Urban Mobility Barriers}
\textbf{(a)} 
Sample of detected mobility barriers in Boston. Lines connect CBG pairs with barriers; CBGs involved in over 100 barriers are highlighted. Some align with visible infrastructure, while others occur between adjacent areas without physical separation.
\textbf{(b)} Normalized likelihood ratio test showing the explanatory power of four factor groups across distances and years. POI factors consistently explain the most variance; physical barriers matter at short distances, while demographic and administrative differences persist across all ranges.
\textbf{(c)} 
Heatmap of logistic regression coefficients for barrier predictors across distances in 11 CBSAs, including POI features, physical infrastructure, demographic differences, and county boundaries. Empty symbols correspond to statistically non-significant coefficients (p-value $< 0.05$). \textbf{(d)} Breakdown of the intervening opportunities effect shown in (c), disaggregated by POI category.
See SI Note~\ref{supp:barrier_composition_by_year} for other years, and SI Note~\ref{supp:barrier_composition_by_city} for each metropolitan area.
}
\label{fig:3}
\end{figure*}

\subsection*{Identification and Nature of Mobility Barriers}
To quantify those mobility barriers, we identify CBG pairs where observed mobility is significantly lower than expected given their geographic distance. Specifically, we compare embedding-based distances to geographic distances (Fig.~\ref{fig:2}d) and define a barrier between CBG $i$ and $j$ when the embedding distance $d^{(e)}_{ij}$ is substantially greater than the value predicted from physical distance $d^{(p)}_{ij}$ using a log-linear model (see Methods). We focus on CBG pairs within 20 km to account for CBSA size and minimize broader spatial effects. Fig.~\ref{fig:3}a shows a sample of barriers under 3 km in Boston. While some align with visible infrastructure, such as rivers or highways, many occur between nearby neighborhoods without a clear physical separation. Some CBGs are also involved in many barriers. To understand the factors associated with barrier presence, we examine four groups of factors known to shape human mobility \cite{jacobs_death_1993}. First, we consider physical barriers measured by the number of highways, railways, parks, and waterways crossed along the direct path between areas $i$ and $j$ \cite{aiello_urban_2024,pinter_neighborhoods_2023,grannis_importance_1998}. Second, we account for amenity structure, using two variables: the number of intervening POIs, which act as alternative destinations between areas $i$ and $j$, and the Jensen–Shannon (JS) distance in POI category composition, capturing functional complementarity between both areas \cite{stouffer1940intervening}. Third, we incorporate sociodemographic differences defined by disparities in income, racial composition, and public transit usage \cite{schelling_models_1969,wang2018urban}. Finally, administrative boundaries \cite{pinter_neighborhoods_2023} are included as a binary variable indicating whether the two CBGs belong to different counties. See SI Note~\ref{supp:distance_measuring} for more details about these groups. These variables are entered into a logistic regression model, both individually and grouped, to evaluate their relationship with the likelihood of a mobility barrier between each CBG pair (see Methods for details).

Our results show that amenity structure is the primary determinant of mobility barriers across all cities, years, and distance ranges, operating through two opposing mechanisms, see Fig.~\ref{fig:3}c and SI Notes~\ref{supp:barrier_composition_by_year} and~\ref{supp:cities_normalzied_flux}. First, the number of intervening opportunities between CBGs is positively associated with barrier formation, as nearby alternative destinations reduce the likelihood of direct movement \cite{jacobs_death_1993,stouffer1940intervening} between areas. In contrast, a greater POI Jensen–Shannon (JS) distance — indicating higher functional complementarity between areas — reduces the likelihood of a barrier, suggesting that people are more likely to travel between areas offering distinct amenity profiles. To unpack the strong overall effect of intervening opportunities seen in panel~\ref{fig:3}c, panel~\ref{fig:3}d breaks down this effect by POI category. Shopping, College, and Sport venues exhibit the most substantial barrier effects at all distances, suggesting they are often final destinations that reduce the likelihood of traveling farther. Categories like Transportation, Service, and Grocery facilitate mobility between short-distance areas, likely because they support ongoing movement. This variation shows that not all amenities create barriers (some support movement between areas), highlighting how the way amenities are arranged in cities can shape mobility and be changed to reduce future barriers. This finding is important not only for planning and intervention but also for mobility modeling: it suggests that the structure of urban venues can either reinforce or mitigate barriers and must be considered explicitly in models that seek to explain or forecast urban movement.

After amenity structure, administrative boundaries are the second strongest predictor of mobility barriers. CBG pairs in different counties are significantly more likely to be barriers, consistent with prior work showing that jurisdictional lines can disrupt movement and neighborhood integration~\cite{pinter_neighborhoods_2023,rinzivillo_discovering_2012}. These invisible divides —shaped by governance, infrastructure, and service differences— limit mobility reinforcing urban fragmentation into counties, cities or districts. Similarly, demographic differences, particularly in income, racial composition, and public transit usage, are strongly associated with the presence of barriers across all cities and distance ranges. This pattern of mobility stratification reflects persistent forms of residential segregation ~\cite{schelling_models_1969} and activity space segregation and isolation in U.S. cities~\cite{wang2018urban,athey_estimating_2021,moro2021mobility},  reinforcing how social and structural divides continue to shape everyday urban movement.

Physical barriers, such as highways, parks, railways, and waterways, are the least predictive of mobility barriers when compared to amenity, administrative, and demographic factors. While these features contribute to restricted movement, particularly at short distances where detours are more costly, their effect diminishes and even reverses at larger spatial scales. For example, highways, which can impose real constraints on mobility at short distances (especially for pedestrians), facilitate movements at longer distances by enabling faster and more direct travel across urban areas. Similarly, parks function as barriers at short distances, contrary to their intended role~\cite{cranz1982politics, fraser2024great}, and may function as border vacuums~\cite{jacobs_death_1993}, limiting mobility between nearby neighborhoods. These results highlight the distance-dependent role of physical infrastructure: features that constrain local movement may enable connectivity at larger scales. 

One possible explanation for the limited contribution of some barriers (e.g., physical) is that they may co-occur with others, making it difficult to isolate their individual effects. However, our analysis of the correlations across barrier types shows only city-wide weak associations (see SI Note \ref{supp:multilinearity}) between them. Thus, although some highways might coincide with administrative boundaries or divide neighborhoods along racial or income lines, these overlaps are not systematic. To assess the relative importance of different types of barriers, we conducted likelihood ratio tests using sequential regressor group ablation (see SI Note~\ref{supp:model_barrier_decompose}). As shown in Fig.~\ref{fig:3}b, amenity-related (POI) variables consistently have the highest explanatory power across all years and distances, confirming their central role in shaping mobility barriers. Physical barriers are most relevant at short distances (under $\lesssim$ 12 km), while their influence declines with distance. Administrative boundaries and demographic differences are particularly important between 5 and 15 km, suggesting that institutional and sociodemographic divisions shape mid-range mobility. In 2020, we observed a moderate shift in the relative importance of these factors, especially a reduction in the role of demographic differences and an increase in the role of administrative boundaries, likely due to pandemic-related disruptions implemented at county or city level. However, by 2021, the pattern closely returns to 2019 levels, suggesting that the barriers detected are not temporary artifacts, but rather reflect stable, deeply rooted behavioral structures in how people move in cities.

\begin{figure*}[t]
\centering
\includegraphics[width=\linewidth]{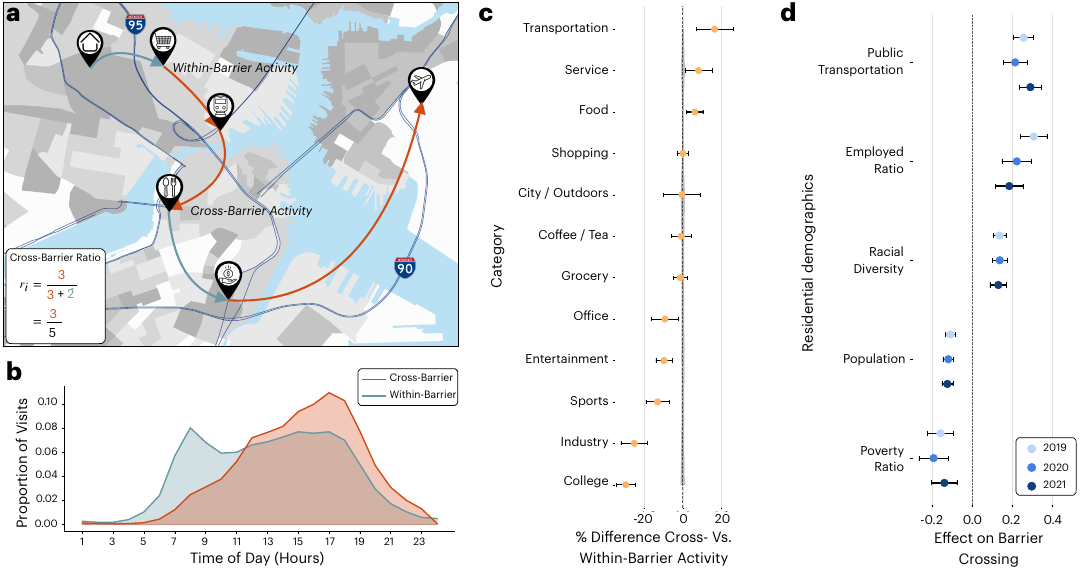}
\caption{
\textbf{Characteristics of cross-barrier mobility patterns.} 
\textbf{(a)} Conceptual framework for classifying activities as cross-barrier (red) or within-barrier (blue) movements. The cross-barrier ratio measures the share of barrier-crossing trips per individual. 
\textbf{(b)} Temporal distribution of cross-barrier activities during the data. Cross-barrier activities peak after 12:00 PM across 11 CBSAs, while within-barrier activities follow a typical commuting pattern.
\textbf{(c)} Percentage difference between activities by category between cross- vs. within-barrier activities. Positive values indicate that those visitation patterns happen more often in cross-barrier movements. 
Error bars represent 95\% confidence intervals calculated across 11 CBSAs. 
\textbf{(d)} Associations between residential demographic characteristics of individuals and cross-barrier ratios from 2019 to 2021. Error bars denote standard errors of coefficient estimates.
}
\label{fig:4}
\end{figure*}

\subsection*{Mobility Across Barriers: Temporal and Social Patterns}
While identifying where barriers exist is essential, understanding who crosses them, when, and why is key to assessing their impact and informing effective interventions. Some barriers are regularly crossed by particular demographic groups for given activities, while others remain restrictive for all. To investigate this, we classify the movements of each individual into two categories: (1) \textit{cross-barrier activities}, where an individual crosses a detected mobility barrier in the city to perform an activity, and (2) \textit{within-barriers activities}, which occur without crossing one (Fig.~\ref{fig:4}a). These types of movements differ markedly. For example, their temporal patterns show a clear contrast: within-barrier activities follow a typical commuting curve, peaking around 8:00 AM, while cross-barrier activities increase steadily throughout the day and peak around 5:00 PM, suggesting they are more discretionary and occur outside standard work hours (Fig.~\ref{fig:4}b).

To examine the types of activities associated with barrier-crossing behavior, we investigate the category of the place visited after every movement (see SI Note~\ref{supp:poi_attribution}). Cross-barrier activities are more frequent than within-barrier ones at transportation-related venues (approximately 20\% higher), services (10\% higher), and food establishments (5\% higher) (Fig.~\ref{fig:4}c). In contrast, individuals are less likely to cross barriers when visiting sports, industry/work, or college-related locations. Furthermore, we find that cross-barrier movements are more exploratory: 67.1\% of them lead to previously unvisited POIs, compared to 58.6\% for within-barrier movements. These patterns suggest again that cross-barrier trips are more often linked to flexible or discretionary activities, indicating that lifestyles based on exploratory behavior may lead to movement across urban barriers \cite{yang_identifying_2023}.

Finally, we examine the residential demographic characteristics of individuals who frequently cross barriers during our 2019-2021 study period. We define the Cross-Barrier Ratio $r_i$ as the proportion of cross-barrier activities for individual $i$ (see Fig.~\ref{fig:4}a and detailed methodology in SI Note~\ref{supp:cbr_definition}). Using linear regression models to predict $r_i$ at the individual level using their home CBG-level demographic information (see SI Note~\ref{supp:cross_barrier_ratio}), we identify key predictors of cross-barrier mobility. As shown in Fig.~\ref{fig:4}d, public transit usage, higher employment rates, and racial diversity are positively associated with more frequent barrier crossing. Associations remain consistent across all three years, indicating that the characteristics of individuals likely to traverse barriers were stable even during the COVID-19 pandemic. Overall, the results indicate that individuals embedded in more connected, transit-accessible, and socially diverse environments are more likely to overcome mobility barriers in their daily lives.

\section*{Discussion}
Cities are not only defined by their physical layout, but how people move through them. While previous research has documented how visible structures, such as highways, rivers, and transit networks, can constrain movement~\cite{van2020missing,beiro2018shopping, cagney2020urban, fraser2024great,pinter_neighborhoods_2023,aiello_urban_2024}, our results show that amenity structure and invisible social and institutional boundaries play a more important role in structuring everyday accessibility ~\cite{dong2020segregated,moro2021mobility,wang2018urban}. For example, the marginal effect of a shopping POI on barrier formation at short distances between areas is about twice as large as that of a highway between them (see SI Fig. \ref{supp_fig:main_results_year}). While intervening opportunities suppress long-distance trips, functional complementarity across neighborhoods increases connectivity. This finding not only reinforces classic urban theories \cite{stouffer1940intervening,jacobs_death_1993}, but also provides new empirical evidence that the urban opportunity landscape is the key behavioral constraint. Importantly, not all amenities act as barriers: transportation and service-related places, for example, tend to promote cross-boundary movement. This highlights a critical planning insight—cities are not only shaped by what exists, but by how it is spatially distributed. The layout of amenities, shaped by zoning, land use, and economic incentives, can either reinforce or reduce barriers. As a result, policies that promote local self-sufficiency, such as the 15-minute city concept, may inadvertently increase fragmentation by limiting exposure to diversity and discouraging cross-neighborhood exploration \cite{xu_using_2025,abbiasov_15-minute_2024}. 

Behavioral patterns around barrier-crossing offer further insight into urban accessibility inequality. We find that discretionary, exploratory trips are associated with movement across barriers, and that individuals who live in neighborhoods with fewer barriers (more demographically diverse, better connected through public transportation and work opportunities) are more likely to cross into other areas \cite{moro2021mobility, yang_identifying_2023}. These patterns are consistent across time—including during pandemic-related disruptions—suggesting that these behavioral boundaries are persistent and deeply embedded in urban life. This opens a window for potential interventions: enabling exploration through targeted investments in transit, public space, or complementary amenities may offer a way to increase mobility, promote spatial integration, and  expand access to opportunity.

Our results also highlight a limitation in widely used models of urban mobility  \cite{zipf1946p,simini2021deep,schlapfer_universal_2021,liang_calibrating_2020} that assume uniform interactions in space governed primarily by distance and population, or by coarse measures of intervening opportunities. While useful for broad predictions, these models overlook key drivers of mobility barriers, such as amenity structure and complementarity, demographic differences, and administrative boundaries. As a result, they may overestimate accessibility and, when used in downstream applications, such as service planning or infrastructure allocation, this may unintentionally reinforce the very barriers they fail to detect.

While our approach offers a detailed and flexible method for identifying mobility barriers, it also has limitations. First, the analysis is correlational and does not establish causal direction. For example, income disparities may reduce mobility by limiting access to transportation, but low mobility could also reinforce inequality by restricting access to jobs, services, or education~\cite{quillian2014does, bor2017population, sharkey2014and}. Second, our model assumes symmetric mobility flows and distances to maintain equivalence with the gravity model~\cite{murray2023unsupervised}. However, real-world flows are often asymmetric: for instance, people may travel more from low-amenity to high-amenity areas than vice versa. Demographic differences may also create directional effects. Future work could adopt directional embedding models~\cite{song2018directional} to capture those asymmetries. Finally, our results might guide the use of natural experiments and spatial causal discovery to explore causal relationships in mobility barriers.

Ultimately, our methodology provides a scalable and adaptable framework to understand what accessibility means in urban systems. Unlike traditional models, it captures the behavioral, social, and institutional constraints that shape real-world movement, complementing previous studies that examine partial aspects of mobility inequalities in urban areas~\cite{wang2018urban,abideen2020deep,pinter2023quantifying,aiello_urban_2024}. Since mobility is at the core of understanding human nature and socioeconomic phenomena, our approach can have important consequences  to other important domains, such as epidemic spreading~\cite{barbosa2018human,chang2021mobility}, innovation~\cite{liang_intercity_2024, atkin_returns_2022}, and economic opportunities~\cite{chetty2016effects,chetty_social_2022}, where physical proximity alone fails to explain patterns of interaction \cite{brockmann_hidden_2013}. 
Neural embedding-based models like ours offer a powerful tool to measure complex functional access, uncover spatial inequality, and inform more inclusive approaches to urban planning and policy. As cities confront rising inequality, the need for frameworks that reflect real accessibility —grounded in behavior rather than geography— becomes increasingly urgent. Access is not defined by meters or minutes, but by the ability to move across the social, economic, and institutional barriers that structure urban life.

\section*{Methods}
\subsection*{Data} \label{materialandmethods:data}
We use anonymized, privacy-enhanced individual-level mobility data from Spectus (\url{https://spectus.ai}) from 25.4 million opted-in users, collected under General Data Protection Regulation (GDPR) and California Consumer Privacy Act (CCPA) compliant frameworks. Our analysis focuses on user visits (stops/stays)\cite{moro2021mobility} across the largest 11 US Core-Based Statistical Areas (CBSAs)~\cite{cbsa} which together include approximately 89.97 million residents (26.3\% of the U.S. population) and span the period from 2019 to 2021, totaling around 6.4 billion visits. CBSAs were selected as the primary geographic unit because they represent functionally integrated metropolitan areas with strong social and economic interconnections and population-dense cores. To minimize seasonal effects, we only analyzed mobility across consecutive 6-month periods from October to March of each consecutive year. All visitation data was up-leveled at the Census Block Group (CBG) level. See SI Note~\ref{supp:stays}) and SI Note~\ref{supp:representativity} for more information about the mobility data and its demographic representativity.

Demographic data at the CBG level were obtained from the ACS 5-year data published from 2019 to 2021 \cite{acs} (SI Note~\ref{supp:demo}). Physical barrier data were obtained from OpenStreetMap~\cite{OpenStreetMap}, utilizing map feature tags to identify polygons representing highways, and parks, and water bodies. Physical barriers between CBGs were quantified by counting the number of obstacles that intersect the direct lines connecting CBG centroids. Detailed specifications for physical barrier classification and measurement are provided in SI Note~\ref{supp:spatial_data} and SI Note~\ref{supp:physical_barrier} respectively. POI data were collected from Foursquare Open Source Places~\cite{foursquare} and SafeGraph Places (via Dewey Data), totaling approximately 2.5 million venues in the 11 CBSAs. Each POI was assigned to one of 20 activity categories using a custom taxonomy (see SI Note~\ref{supp:poi} for more information).

\subsection*{Word2Vec Embeddings}  \label{materialandmethods:methods}
We represent individual mobility by treating locations as words, $c_t$, and individual trajectories as sentences, denoted by $(c_{1}, c_{2}, \dots, c_{T})$, where each $c_{t}$ is a CBG visited by a given user at time $t$ (see SI Note~\ref{supp:traj} for details). These CBG-based user trajectories are used as input to the standard skip-gram negative sampling Word2Vec model~\cite{mikolov2013distributed, mikolov2013efficient}. The model learns a dense vector $\mathbf{w}_i$ for each CBG $i$, such that the distance between vectors reflects the likelihood of co-occurrence in similar movement contexts. Formally, given a context window size $w$, the model learns the probability $p(c_{t+\tau} \mid c_t)$ from context $-w \le \tau \le w$, $\tau \ne 0$, by maximizing the following log-likelihood:$
\mathcal{J} = \frac{1}{T}\sum_{t=1}^T \sum_{\substack{-w \le \tau \le w \ \tau \ne 0}} \log p(c_{t+\tau} \mid c_t) $
where $ p(j|i) = \exp (\mathbf{u_{j}} \cdot \mathbf{v_{i}}) / Z_i$. 
Here, $\mathbf{v}$ and $\mathbf{u}$ are the ``in-vector'' and ``out-vector'' respectively, and $Z_{i}=\sum_{j^\prime \in \mathcal{C}} \exp(\mathbf{u_{j^\prime} \cdot \mathbf{v_{i}}})$ is the normalization constant, where $\mathcal{C}$ is the set of all CBGs. Following the standard practice~\cite{murray2023unsupervised,tshitoyan2019unsupervised,garg2018word}, we use the ``in-vectors'' $\mathbf{w} = \mathbf{v}$ for further analysis due to their superior ability in capturing semantic relationships. Details of training parameters for the Word2Vec  model can be found in SI Note~\ref{supp:w2v}.


\subsection*{Defining Mobility Barriers}\label{materialandmethods:detecting_barriers}
We define the set of mobility barriers in an urban area, $\mathbb{B}$, as pairs of CBGs where the observed embedding (cosine) distance is significantly greater than expected based on their geographic distance. As shown in Fig.~\ref{fig:2}d, we model the expected embedding distance $\hat{d}_{ij}^{(e)}$ as a log-linear function of geographic distance: $\hat{d}_{ij}^{(e)} \sim \beta \log d_{ij}^{(p)} + \epsilon_{ij}$,
where $\epsilon_{ij}$ is the error term. The residual $r_{ij} = d_{ij}^{(e)} - \hat{d}_{ij}^{(e)}$ measures the extent to which the observed embedding distance exceeds the expected value. 
We restrict the analysis to CBG pairs within 20 km and group them into 1 km distance bins. Within each bin, we identify mobility barriers as the top 5\% of pairs with the largest positive residuals. This approach systematically detects local mobility barriers where observed movement is notably less than predicted by spatial proximity.

\subsection*{Logistic Regression Model}
To understand the nature of mobility barriers, we design a conditional logistic regression model:
\begin{equation}\label{eq:logistic}
    \text{logit}\ P(i\leftrightarrow j \in \mathbb{B} | d_{ij}^{(p)}) \sim 
    \{d_{ij,\text{POI}}\} + 
    \{d_{ij,\text{Phy}}\} + 
    \{d_{ij,\text{Demo}}\} +  
    \{d_{ij,\text{County}}\}
\end{equation}
where the log-odds of pair of CBGs $i$ and $j$ being barrier ($i\leftrightarrow j \in \mathbb{B}$) at a given physical distance $d_{ij}^{(p)}$ are modeled as a linear combination of POI distance variables $\{d_{\text{POI}}\}$, physical barrier variables $\{d_{\text{Phy}}\}$, demographic distance variables $\{d_{\text{Demo}}\}$, and administrative boundary differences $\{d_{\text{County}}\}$ between them (see SI Note~\ref{supp:model_barrier_decompose} for more details).
To train the model, the positive class consists of CBG pairs identified as barriers within each 1 km geographical distance $d_{ij}^{(p)}$ group. The negative class is randomly sampled from all CBG pairs in the same distance group, balanced in size to the positive class.
All distance variables are standardized within each group. This modeling framework enables the interpretation of the relative importance of each regressor by comparing their coefficients. 
Besides, the model is applied separately within each geographical distance group to control for spatial effects. Full results for different years, cities, and specifications of the model, along with robustness checks of the model results can be found in SI Notes \ref{supp:threshold_robustness}, \ref{supp:h3_robustness}, \ref{supp:barrier_composition_by_year}, and \ref{supp:barrier_composition_by_city}.

\section*{Data Availability}
The data supporting the findings of this study are available from Spectus through their Social Impact program; however, restrictions apply to the availability of these data, which were used under the license for the current study and are therefore not publicly available. 
Information on how to request access to the data, its conditions, and limitations can be found at \url{https://spectus.ai/social-impact/}. 
Data about the POI locations was provided by Safegraph and Foursquare. The Foursquare data is publicly available at \url{https://docs.foursquare.com/data-products/docs/access-fsq-os-places}. Safegraph POI dataset is available through the Dewey platform \url{https://app.deweydata.io/home}. Census data (American Community Survey tables and tiger shapefiles) can be downloaded from the US Census Bureau via \url{https://www.census.gov}. A description of these datasets is given in SI Note~\ref{supp:data}.

\section*{Code Availability}
The analysis was conducted using Python and R. The code to reproduce the main results in the figures from the aggregated data is publicly available on GitHub via \url{https://github.com/SUNLab-NetSI/invisible_barriers}. See SI Note~\ref{supp:library} for more information.

\section*{Acknowledgments} 
We thank Hamish Gibbs, Saumitra Kulkarni, and Bijin Joseph, Luca Aiello, and Dakota Murray for their valuable insights on the paper. E.M. and G.W. acknowledge support from the U.S. National Science Foundation under Grants 2420945 and 2427150. 

\section*{Author contributions statement}
G.W., M.K., and E.M. performed research; All authors supervised research, analyzed the results, and co-wrote the manuscript.

\section*{Competing interests} 
The authors declare no competing interests. 

\clearpage
\section*{Supplementary Information}

\section{Data}\label{supp:data}
\subsection{Locational Data}
\subsubsection{Stays}\label{supp:stays}
We obtained mobility data from Spectus, a location intelligence and measurement company. 
The dataset comprises anonymized GPS records from users who opted in to share data anonymously in U.S. metropolitan areas over three semiannual periods (October–March) spanning 2019–2022, following the methods in~\cite{moro2021mobility}. 
Data access was granted under Spectus’s Data for Good program, which provides de-identified, privacy-enhanced mobility data exclusively for academic research and humanitarian initiatives. 
All researchers were contractually prohibited from redistributing the data or attempting re-identification. 
Mobility data is derived from users who opted in to share their data anonymously through a General Data Protection Regulation (GDPR) and California Consumer Privacy Act (CCPA) compliant framework.

Data consists of users/devices stay events, denoted as \(\{\mathbf{x}_{i,t}\}\), where an anonymous user \(i\) remained stationary at location $\mathbf{x}$ at time \(t\). It also contains information their home Census Block Group (CBG), estimated from the stays at night. To generate individual trajectories for our Word2Vec model, we up-leveled each stay to the Census Block Group where it happened $\mathbf{x}_{i,t} \rightarrow {c}_{i,t}$. For this study, we use stays filtered to include only users whose home CBGs fall within the relevant Core Based Statistical Area (CBSA). 

To ensure consistency and comparability across years, we focus exclusively on the October–March periods for 2019–2022. This choice reduces the influence of seasonally atypical behaviors associated with summer months, such as tourism or vacations, and excludes the onset of COVID-19 lockdowns that began in mid-March 2020. This window allows us to capture more stable patterns of routine urban mobility when behavioral barriers are more likely to reflect persistent structural and social factors rather than short-term disruptions or seasonal variation.

Summary statistics for each time period are shown in Tables~\ref{tbl:stats_2019},~\ref{tbl:stats_2020}, and~\ref{tbl:stats_2021}.

\begin{table}[!htb]\centering
\caption{Description of each of the core-based statistical areas considered and some statistics about our data set from October 2019 to March 2020.}
\label{tbl:stats_2019}
\begin{tabular}{lllll}
\toprule
CBSA Name & State & Population & \# devices & \# stays \\
\midrule
New York-Newark-Jersey City & NY-NJ-PA & 19.97 M & 2239.06 K & 540.52 M \\
Los Angeles-Long Beach-Anaheim & CA & 13.25 M & 1359.86 K & 337.52 M \\
Chicago-Naperville-Elgin & IL-IN-WI & 9.51 M & 1343.08 K & 320.03 M \\
Dallas-Fort Worth-Arlington & TX & 7.39 M & 1328.85 K & 362.11 M \\
Washington-Arlington-Alexandria & DC-VA-MD-WV & 6.20 M & 675.06 K & 151.07 M \\
Miami-Fort Lauderdale-Pompano Beach & FL & 6.09 M & 1017.91 K & 242.28 M \\
Philadelphia-Camden-Wilmington & PA-NJ-DE-MD & 6.08 M & 727.76 K & 175.99 M \\
Boston-Cambridge-Newton & MA-NH & 4.83 M & 466.72 K & 104.26 M \\
Detroit-Warren-Dearborn & MI & 4.32 M & 634.00 K & 155.49 M \\
Seattle-Tacoma-Bellevue & WA & 3.87 M & 437.13 K & 101.68 M \\
San Francisco-Oakland-Berkeley & CA & 3.83 M & 366.46 K & 78.78 M \\
Total &  & 85.33 M & 10595.89 K & 2569.75 M \\
\bottomrule
\end{tabular}

\end{table}

\begin{table}[!htb]\centering
\caption{Description of each of the core-based statistical areas considered and some statistics about our data set from October 2020 to March 2021.}
\label{tbl:stats_2020}
\begin{tabular}{lllll}
\toprule
CBSA Name & State & Population & \# devices & \# stays \\
\midrule
New York-Newark-Jersey City & NY-NJ-PA & 19.94 M & 1587.51 K & 331.99 M \\
Los Angeles-Long Beach-Anaheim & CA & 13.21 M & 885.00 K & 188.38 M \\
Chicago-Naperville-Elgin & IL-IN-WI & 9.48 M & 973.35 K & 207.25 M \\
Dallas-Fort Worth-Arlington & TX & 7.52 M & 1059.79 K & 266.44 M \\
Washington-Arlington-Alexandria & DC-VA-MD-WV & 6.25 M & 469.13 K & 87.03 M \\
Miami-Fort Lauderdale-Pompano Beach & FL & 6.13 M & 716.74 K & 159.26 M \\
Philadelphia-Camden-Wilmington & PA-NJ-DE-MD & 6.09 M & 541.56 K & 112.77 M \\
Boston-Cambridge-Newton & MA-NH & 4.85 M & 325.92 K & 64.96 M \\
Detroit-Warren-Dearborn & MI & 4.32 M & 464.20 K & 101.18 M \\
Seattle-Tacoma-Bellevue & WA & 3.93 M & 304.27 K & 60.42 M \\
San Francisco-Oakland-Berkeley & CA & 3.83 M & 245.42 K & 43.24 M \\
Total &  & 85.56 M & 7572.89 K & 1622.92 M \\
\bottomrule
\end{tabular}

\end{table}

\begin{table}[!htb]\centering
\caption{Description of each of the core-based statistical areas considered and some statistics about our data set from October 2021 to March 2022.}
\label{tbl:stats_2021}
\begin{tabular}{lllll}
\toprule
CBSA Name & State & Population & \# devices & \# stays \\
\midrule
New York-Newark-Jersey City & NY-NJ-PA & 20.71 M & 1556.83 K & 468.15 M \\
Los Angeles-Long Beach-Anaheim & CA & 13.20 M & 836.27 K & 262.34 M \\
Chicago-Naperville-Elgin & IL-IN-WI & 9.61 M & 853.14 K & 269.31 M \\
Dallas-Fort Worth-Arlington & TX & 7.61 M & 1050.64 K & 353.89 M \\
Washington-Arlington-Alexandria & DC-VA-MD-WV & 6.33 M & 452.01 K & 123.72 M \\
Philadelphia-Camden-Wilmington & PA-NJ-DE-MD & 6.22 M & 510.27 K & 156.91 M \\
Miami-Fort Lauderdale-Pompano Beach & FL & 6.11 M & 724.98 K & 212.56 M \\
Boston-Cambridge-Newton & MA-NH & 4.91 M & 317.05 K & 91.12 M \\
Detroit-Warren-Dearborn & MI & 4.38 M & 402.66 K & 125.52 M \\
Seattle-Tacoma-Bellevue & WA & 3.97 M & 275.74 K & 79.74 M \\
San Francisco-Oakland-Berkeley & CA & 3.86 M & 222.87 K & 59.58 M \\
Total &  & 86.91 M & 7202.46 K & 2202.84 M \\
\bottomrule
\end{tabular}

\end{table}

\subsubsection{Trajectories}\label{supp:traj}
The input corpus for the Word2Vec (Skip-gram with Negative Sampling, SGNS) model consists of sequences analogous to sentences in natural language processing, which in our study correspond to human mobility trajectories from opted-in users. We construct these sequences by aggregating individual stays at the CBG level in chronological order, where each sequence represents the complete set of CBGs visited by a single user during the study period without additional temporal separators.

Formally, a trajectory can be denoted as $\mathbf{T}_i = (c_{i,1}, c_{i,2}, \ldots, c_{i,T_i})$, where $c_{i,t}$ represents the $t$-th CBG visited in the trajectory of anonymous user $i$, and $T_i$ denotes the total number of CBG visits for user $i$. 
The temporal context is preserved through the sequential ordering, where $c_{i,t+1}$ and $c_{i,t-1}$ represent the CBGs visited immediately after and before $c_{i,t}$, respectively. 
Since users may visit multiple points of interest (POI) within a single CBG, consecutive identical CBG visits can occur, i.e., $c_{i,t} = c_{i,t-1}$.

Due to inherent limitations in mobile device sensing capabilities, not all user visits are accurately captured in mobility datasets. This results in incomplete trajectories that may yield spurious transition patterns. To address these data quality issues and recover more accurate mobility patterns, we implement the following preprocessing procedures:
\begin{enumerate}
    \item \textbf{Minimum activity threshold:} Users with fewer than 5 recorded stays during the entire study period (6 months) are excluded from the training corpus to ensure long enough mobility signal.
    \item \textbf{Daily mobility filtering:} For any given day, if a user has fewer than 2 stays recorded on that day, the corresponding trajectory segment is removed as it lacks meaningful inter-CBG mobility information.
    \item \textbf{Temporal gap handling:} If the temporal gap between the end of one stay and the beginning of the subsequent stay exceeds $\Delta_t = 1$ hour, the trajectory segment for that date is considered incomplete and excluded from analysis.
    \item \textbf{Spatial proximity merging:} When consecutive stays occur within a distance threshold of $\Delta_d = 50$ meters, they are merged and treated as a single extended visit to account for GPS noise and minor location variations.
\end{enumerate}

\subsubsection{Trajectory pruning}\label{supp:flow_pruning}
Based on the pre-processed trajectories described in Section~\ref{supp:traj}, we applied two thresholds for trajectory pruning to address overfitting issues in the Word2Vec training process. 

\begin{figure}[!h]
    \centering
    \includegraphics[width=\linewidth]{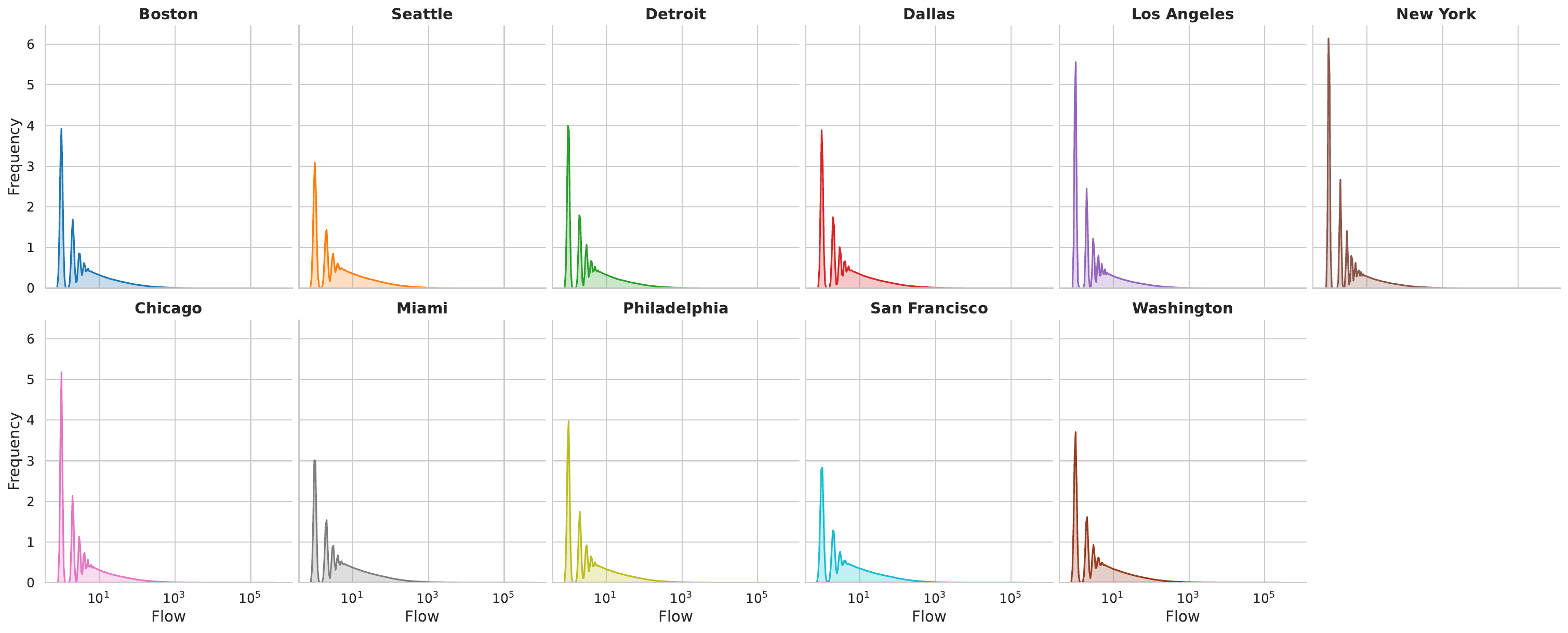}
    \caption[Distribution of raw flows for each CBSA in 2019.]{Distribution of raw flows for each CBSA in 2019. Flow values are displayed on a logarithmic scale along the x-axis.}
    \label{supp_fig:before_flow_distribution_2019}
\end{figure}

The Word2Vec model is designed to predict neighboring entities $\{c_{i,t-n}, \ldots, c_{i,t+n}\}$ given a center entity $c_{i,t}$ in the training corpus, where $n$ is the window size of the neighborhood.
In the context of training on human trajectory data, the model's objective is to learn inter-CBG transition patterns. 
However, due to the shallow architecture of Word2Vec~\cite{mikolov2013distributed, mikolov2013efficient} and the inherent sparsity of mobility data (see Fig.~\ref{supp_fig:before_flow_distribution_2019}), the model is prone to overfitting.

The overfitting in our study occurs in two ways due to the highly skewed distribution of mobility flows in urban areas. First, we observe a high density of rare flows, i.e. very small flows between CBGs. Since there are few co-occurrences of such flow patterns, the model attempts to memorize these cases, resulting in embedding distances that are disproportionately small relative to the actual mobility relationships. Second, we encounter extremely high-frequency flows caused by super-popular destinations such as airports and railway stations. Due to the exceptionally high weights of these flows in the training corpus, the model becomes biased toward these patterns, distorting the embedding distances to other entities.

\begin{figure}[!h]
    \centering
    \includegraphics[width=\linewidth]{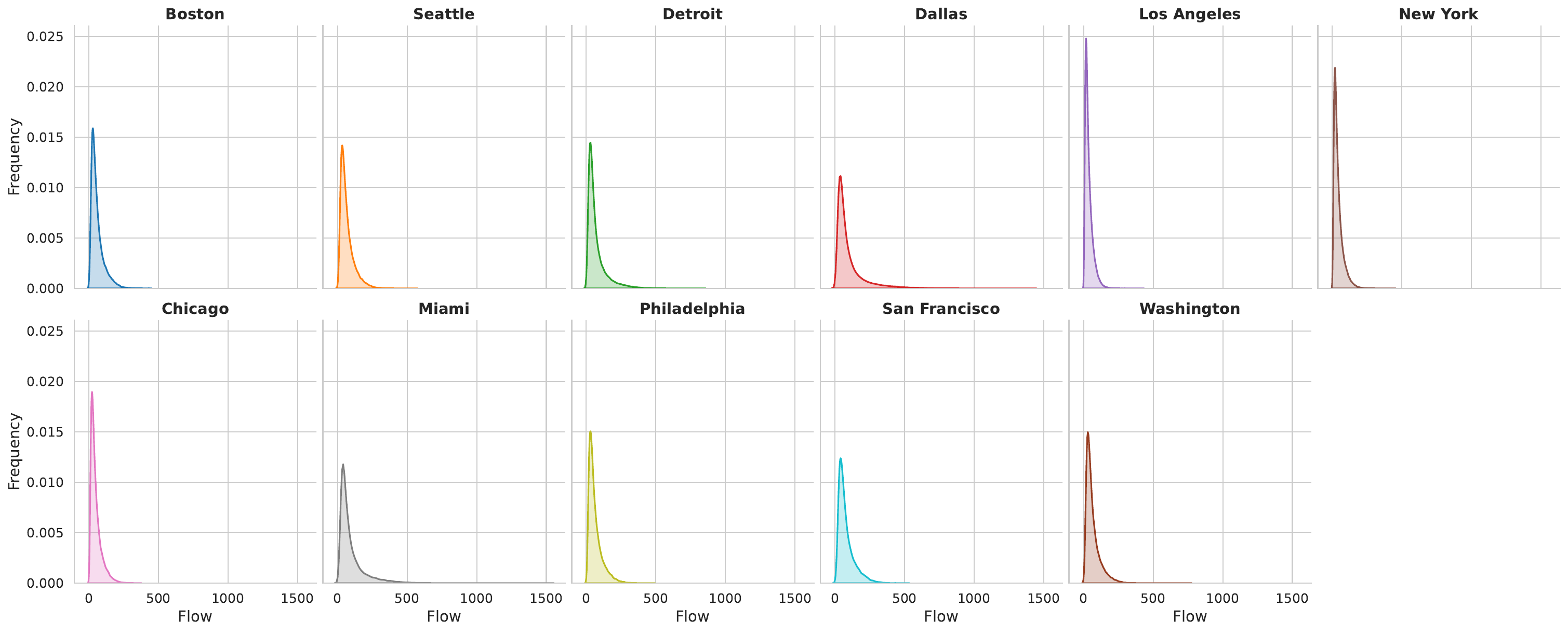}
    \caption[Distribution of pruned flows for each CBSA in 2019.]{Distribution of pruned flows for each CBSA in 2019.}
    \label{supp_fig:after_flow_distribution_2019}
\end{figure}

To mitigate these issues, we constrain flows to lie within the quantile range $[0.95, 0.995]$. When a flow lies outside this range, we remove the entire trajectory segment for that date rather than removing individual flows, thereby avoiding the creation of artificial flow patterns in the training corpus.  The processed flow distributions can be found in Fig.~\ref{supp_fig:after_flow_distribution_2019}. Since the exclusion of high-flow CBG pairs from our model creates artificial mobility barriers between these locations in subsequent analysis, we exclude such pruned CBG pairs from our barrier detection method. Previous work~\cite{lauren2021constraining} on language texts has demonstrated that constraining frequency weights achieves comparable results in word similarity and sequence labeling tasks, but reduces the presence of outliers. In fact, our results are not affected by this pruning, as we can see through extensive robustness checks in Section~\ref{supp:threshold_robustness}. 

\subsection{Representativeness of the Locational data}\label{supp:representativity}
Our location data is derived from smartphone devices across major U.S. metropolitan areas. 
While smartphone penetration rates are high in urban populations, the representativeness of our sample of 25.4 million devices relative to the underlying demographic distribution requires empirical validation.

\begin{figure}[t]
    \centering
    \includegraphics[width=\linewidth]{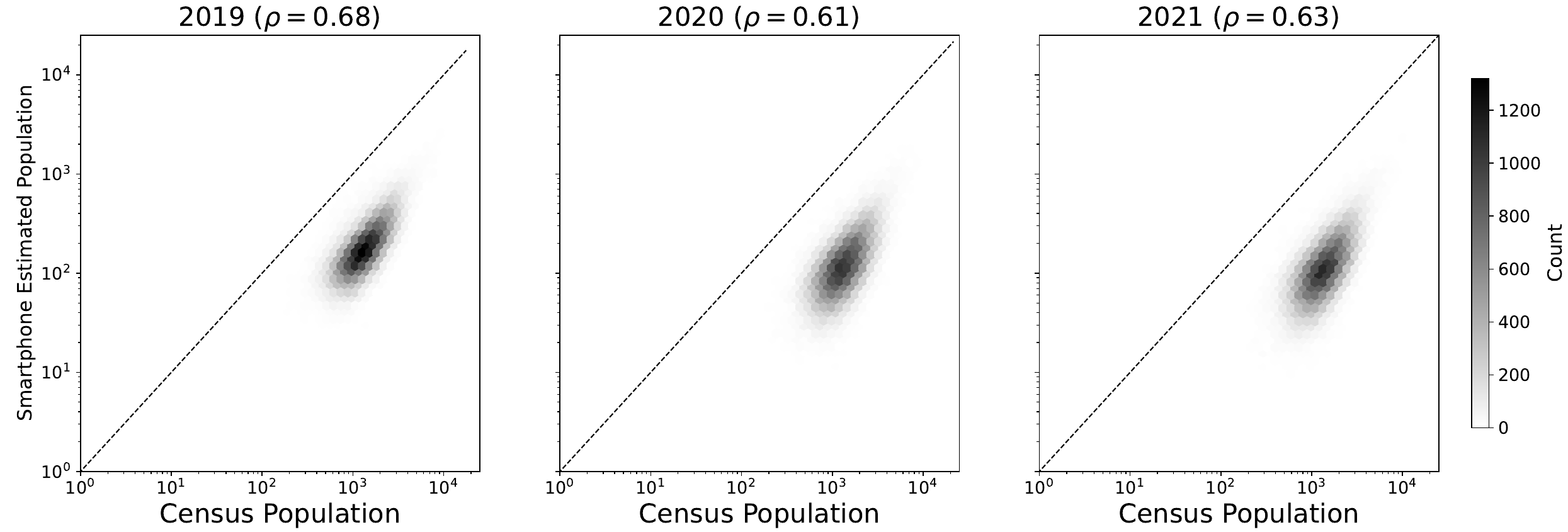}
    \caption[Correlation between the smartphone population detected in our data and Census population.]{Correlation between the smartphone population detected in our data and Census population. The Spearman correlation between them is shown at the top.}
    \label{supp_fig:representative}
\end{figure}

To evaluate representativeness, we compare the spatial distribution of devices in our dataset against population estimates from the American Community Survey (ACS) 5-year estimates spanning 2019-2021 at the CBG level across all 11 metropolitan areas in our study. 
For the 2020 and 2021 periods, we employed reweighted demographic estimates based on the corresponding ACS 5-year data to account for data limitations. Detailed methodology for demographic reweighting is provided in Section~\ref{supp:demo}.
Fig.~\ref{supp_fig:representative} presents this comparison, revealing a moderately strong positive correlation between observed device density and official population counts ($\rho = 0.64 \pm 0.02$ on average across three years).

\subsection{Demographic data} \label{supp:demo}
Demographic data at the CBG level was obtained from the American Community Survey (ACS) 5-year data published between 2019 to 2021~\cite{acs}. Each demographic characteristic corresponds to a specific ACS data table. In our study, we used the following demographic characteristics \emph{estimates} at the CBG level:
\begin{enumerate}
    \item Proportion of population by ethnicity (Table B03002); 
    \item Median value of owner-occupied housing units (Table B25077); 
    \item Proportion of workers commuting by public transportation (Table B08301);
    \item Proportion of employed population (Table B23025);
    \item Proportion of people with a college education or beyond (Table B15002);
    \item Proportion of family household type (Table B11016); and
    \item Median household income in the past 12 months (Table B19013).
\end{enumerate}


\subsection{POI data}\label{supp:poi}
\begin{figure}[!tb]
    \centering
    \includegraphics[width=\linewidth]{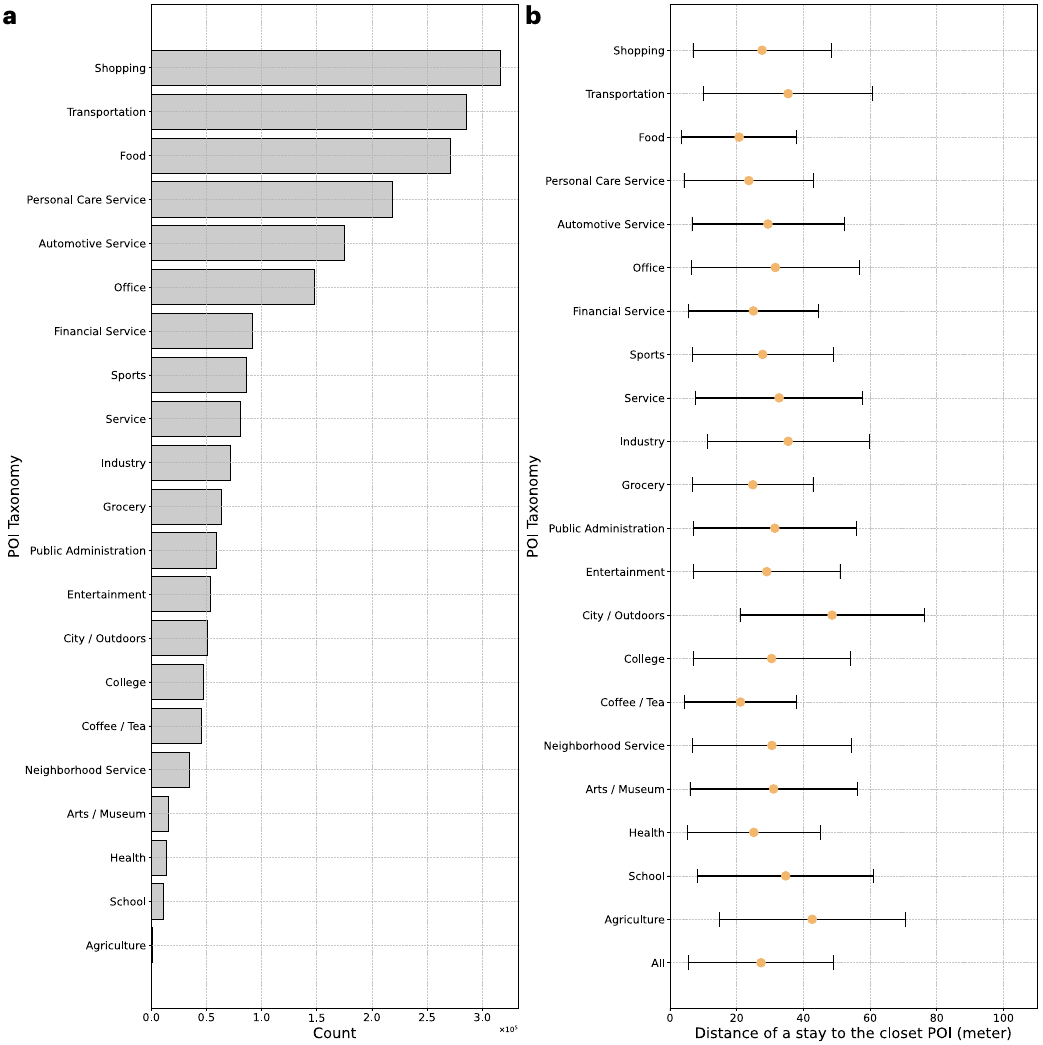}
    \caption[Distribution of POI in 11 CBSA by category and distance from stays to those POIs.]{
    \textbf{(a)} Distribution of points of interest from 11 CBSAs across categories in the venue dataset. 
    \textbf{(b)} Distribution of spatial attribution distances across venue categories for 2019. Error bars denote standard errors of attribution distances.
    }
    \label{supp_fig:poi_dist}
\end{figure}

Our Point of Interest (POI) data contains information about the precise locations of venues and their corresponding activity category. We construct the POI data set from two complementary datasets to ensure comprehensive coverage in urban areas: Foursquare Open Source Places~\cite{foursquare} and SafeGraph Places Data, accessed through the Dewey Data platform (\url{https://www.deweydata.io/}) under SafeGraph's standard usage terms and conditions (\url{https://www.safegraph.com/}). 
The integrated dataset comprises approximately 2.5 million POIs distributed across 11 CBSAs. To facilitate systematic analysis, we developed a custom taxonomy that categorizes these venues into 20 distinct classifications, as illustrated in Fig.~\ref{supp_fig:poi_dist}a. This taxonomy enables consistent POI categorization and supports comparative analysis across different location types and urban contexts. 

\subsubsection{Spatial Attribution of Stays to Venues}\label{supp:poi_attribution}
To identity the activity behind any stay, we attributed each of them to a particular POI. Following established methodologies in location inference from sparse mobile data~\cite{cuttone2014inferring,moro2021mobility}, we employ a proximity-based attribution approach where each stay is assigned to its nearest POI within the dataset. Both stays and POIs are represented as discrete geographic points. 
The attribution process implements a nearest-neighbor algorithm with a maximum distance constraint of $d_{\text{max}} = 100$ meters to prevent erroneous associations with distant venues. 
Stays that fall beyond this threshold from any venue remain unattributed, ensuring spatial accuracy in our venue-stay associations.
This methodology demonstrates robust performance characteristics: the median distance between attributed stays and their corresponding venues is approximately $20.6$ meters, indicating high spatial precision in the attribution process. 
The distribution of stay-to-venue distances is consistent across venue taxonomies, as presented in Fig.~\ref{supp_fig:poi_dist}b.

\subsection{Spatial Data}\label{supp:spatial_data}
We obtained physical barrier spatial data for all 11 CBSAs examined in this study from the open-source, crowd-sourced platform OpenStreetMap (OSM)~\cite{OpenStreetMap} using the \texttt{OSMnx} Python package v1.9.4~\cite{boeing2025modeling}. 
Specifically, we employed the \path{osmnx.features_from_polygon} function with CBSA polygons obtained from TIGER shapefiles provided by the U.S. Census Bureau~\cite{tiger}, OSM feature tags detailed in Table~\ref{tbl:osm}, and all other default parameters.

The complete catalog of OSM features and tags is available at \url{https://wiki.openstreetmap.org/wiki/Map_features}. 
The methodology for determining physical barriers between CBG pairs is described in Section~\ref{supp:physical_barrier}.

\begin{table}[!htb]
\centering
\caption{OSM tags grouped by physical barrier type and key.}
\label{tbl:osm}
\begin{tabular}{llp{8cm}}
\toprule
Physical barrier type & OSM Feature (key) & Tags \\
\midrule
Highway & highway & motorway, trunk \\ 
\midrule
Railway & railway & rail, construction, funicular, light\_rail, monorail \\
\midrule
\multirow{4}{*}{Park} & leisure & park, playground, golf\_course, nature\_reserve, recreation\_ground, track, stadium \\ 
 & boundary & forest, national\_park, protected\_area \\ 
 & landuse & forest \\ 
 & natural & wood, heath, grassland, scrub, wetland \\ 
\midrule
\multirow{2}{*}{Waterway} & water & river, oxbow, canal, lake, reservoir, lagoon, moat, wastewater, pond \\
 & place & sea, ocean \\
\bottomrule
\end{tabular}

\end{table}

\section{Visible and invisible barrier metrics}\label{supp:distance_measuring}
\subsection{Measuring POI Intervening Opportunities}
Intervening opportunities represent alternative destinations or attractions that may divert individuals from their originally intended destination during spatial movement. 
Stouffer's law of intervening opportunities~\cite{stouffer1940intervening} formalizes this concept, stating that ``the number of persons going a given distance is directly proportional to the number of opportunities at that distance and inversely proportional to the number of intervening opportunities.''
This principle captures the fundamental trade-off between destination attractiveness and the availability of alternative opportunities along the travel path.

\begin{figure}[tb]
    \centering
    \includegraphics[width=0.7\linewidth]{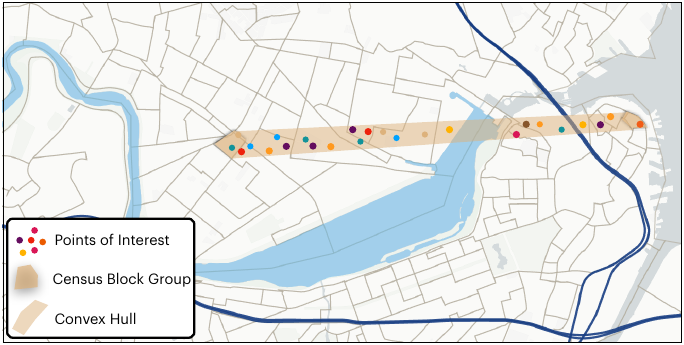}
    \caption[Illustration of POI intervening opportunity quantification for a given CBG pair]{
    Illustration of POI intervening opportunity quantification for a given CBG pair. 
    The convex hull encompasses all POIs that could serve as intervening opportunities between the two CBGs.
    Map was produced in kepler.gl using the TIGER shapeﬁles from the U.S. Census Bureau~\cite{tiger}.
    }
    \label{supp_fig:poi_inter}
\end{figure}

In our analysis, we quantify POI intervening opportunity as the total count of POIs located within the convex hull formed by a pair of CBGs inclusively, as illustrated in Fig.~\ref{supp_fig:poi_inter}.
This geometric approach estimates the potential destinations that could reasonably divert movement between the origin and destination CBG, as well as those in the potential travel corridor.
To prevent multicollinearity problems with ``\# Park Crossed'' in the model given by Eq.~\eqref{eq:logistic}, POIs wtih category of ``City / Outdoors'' are excluded from the POI Intervening Opportunity.

\subsection{Measuring Physical Barriers}\label{supp:physical_barrier}
\begin{figure}[tb]
    \centering
    \includegraphics[width=0.7\linewidth]{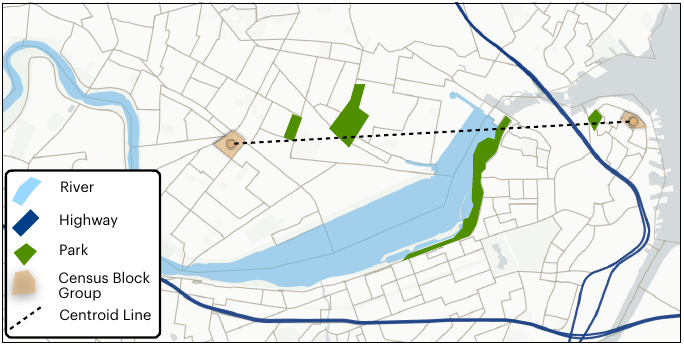}
    \caption[Illustration of physical barrier quantification for a given CBG pair.]{
    Illustration of physical barrier quantification for a given CBG pair.
    The straight line connecting CBG centroids intersects with one river, one highway, and four parks. Map is produced in kepler.gl using the TIGER shapefiles from the U.S. Census Bureau~\cite{tiger}.
    }
    \label{supp_fig:physical_barrier}
\end{figure}

Physical barriers represent spatial elements that potentially impede inter-CBG mobility.
Following established urban studies literature~\cite{pinter2023quantifying, aiello_urban_2024}, we study:
\begin{itemize}
    \item Natural or built infrastructure elements that actively restrict or prevent movement across space, including rivers, highways, and railways ~\cite{pinter2023quantifying, aiello_urban_2024}. These barriers create tangible impediments to pedestrian and vehicular mobility by requiring detours or specialized crossing infrastructure.
    \item Spatial features that unintentionally isolate communities by creating areas of prolonged activity or residence, conceptualized as ``border vacuums"~\cite{jacobs_death_1993}. These include parks, transportation hubs, and large institutional complexes where individuals often spend extended periods.
\end{itemize}
In our analysis, we quantify physical barriers between CBGs as the total count of barriers that intersects with the straight-line connection between the centroids of each CBG pair, as illustrated in Fig.~\ref{supp_fig:physical_barrier}.

\subsection{Measuring Demographic distances}
Three of variables (proportion of population by ethnicity, median value of owner-occupied housing units, and proportion of workers commuting by public transportation) were used to calculate demographic distances (see Section~\ref{supp:distance}) as predictors in our logistic regression model to explain barrier composition patterns (see Section~\ref{supp:model_barrier_decompose}). We also tested distances based on other demographic characteristics; however, these were excluded from the final analysis due to multicollinearity with those three considered. Racial distance was measured using the Jensen-Shannon distance between the distributions of population by ethnicity of different CBG pairs. In particular, if $R_i$ is the distribution of population by ethnicity in CBG $i$, then we get
$$
Race\ Distance_{ij} = \sqrt{\frac{\text{KL}(R_i \parallel M_{ij}) + \text{KL}(R_j \parallel R_j)}{2}}, 
$$
where $M_{ij} = (R_i+R_j)/2$ is the point-wise mean of $R_i$ and $R_j$ and $KL(\cdot || \cdot)$ denote the Kullback-Leibler divergence. Income and Public Transportation distance were calculated as the difference of median value of owner-occupied housing units, and proportion of workers commuting by public transportation between CBG pairs.

\subsection{Embedding Distance Measuring} \label{supp:distance}
In our study, we calculate and use various distances in two contexts: the gravity law of mobility and the logistic regression model.
For the gravity law of mobility, we compute two primary distance measures between CBG pairs.
The geographical distance $d_{ij}^{(p)}$ is calculated as the pairwise great-circle distance between the centroids of CBGs $i$ and $j$.
The embedding distance $d_{ij}$ employs cosine distance to quantify similarity in the learned feature space:
\begin{equation} 
    d_{ij} = 1 - \frac{\mathbf{v}_i \cdot \mathbf{v}_j}{\lVert\mathbf{v}_i\rVert \lVert\mathbf{v}_j\rVert}, 
\end{equation} 
where $\mathbf{v}_i$ and $\mathbf{v}_j$ represent the embedding vectors for CBGs $i$ and $j$, respectively. 

\subsection{Distribution, correlation, and multicolinearity of barrier metrics}\label{supp:multilinearity}
\begin{figure}[htb]
    \centering
    \includegraphics[width=\linewidth]{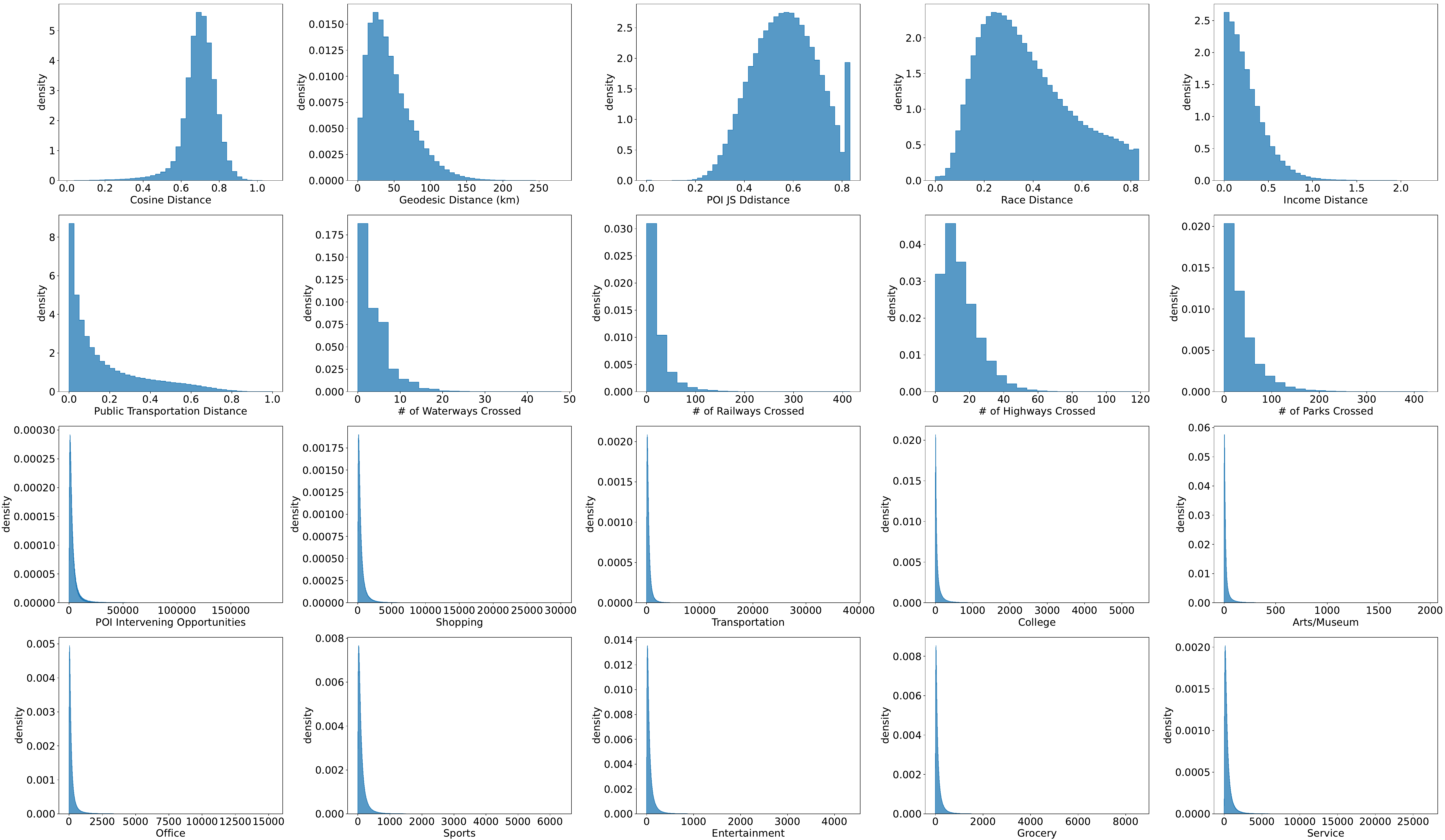}
    \caption[Distribution of distance measures used to characterize barriers across all 11 CBSAs.]{
    Distribution of distance measures used to characterize barriers across all 11 CBSAs.
    }
    \label{supp_fig:data_distribution}
\end{figure}

The distributions of all distance measures can be found in Fig.~\ref{supp_fig:data_distribution}. Since urban structure conditions where some of these barriers might happen, we carefully investigate the correlation between these variables and their multicollinearity. High correlation among our metrics might signal co-occurrence of those barriers that can propagate to multicollinearity and inflate the variance of coefficient estimates in regression models, potentially obscuring the contribution of individual predictors.

\begin{figure}
    \centering
    \includegraphics[width=0.95\linewidth]{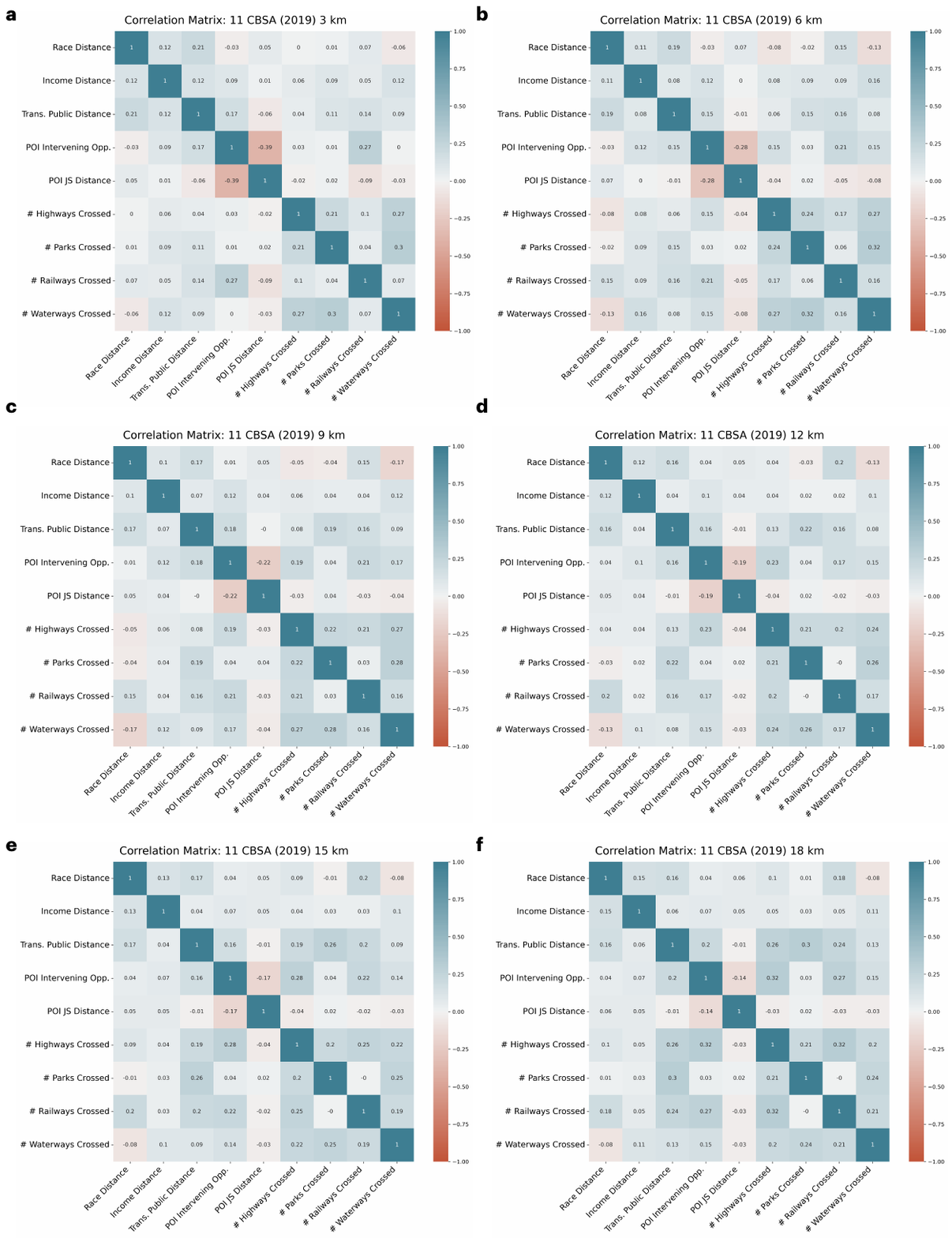}
    \caption[Pairwise correlation matrices for all barrier predictors in six representative distance groups (11 CBSA, 2019).]{
    Pairwise correlation matrices for all predictors in six representative distance groups (11 CBSA, 2019).  
    With the exception of a handful of correlations above 0.30 at the shortest ranges, interdependencies are weak, indicating negligible multicollinearity.
    }
    \label{supp_fig:multicollinearity}
\end{figure}

However, as we can see in (Fig.~\ref{supp_fig:multicollinearity}, across all distance bins, correlations are generally low: apart from the moderate association between \textit{POI Intervening Opportunities} and \textit{POI JS Distance} at the shortest ranges (\,\(\rho \lesssim 0.5\)\,), no coefficient exceeds $\rho \leq 0.35$.  
Because each distance cohort is estimated independently and the few elevated correlations are isolated to specific short‐range models, we retain the full set of regressors in the final specification.  This choice preserves comparability across distance bands while posing minimal risk of variance inflation.

\section{Cross Barrier Ratio}\label{supp:cbr_definition}
\begin{figure}[t]
    \centering
    \includegraphics[width=\linewidth]{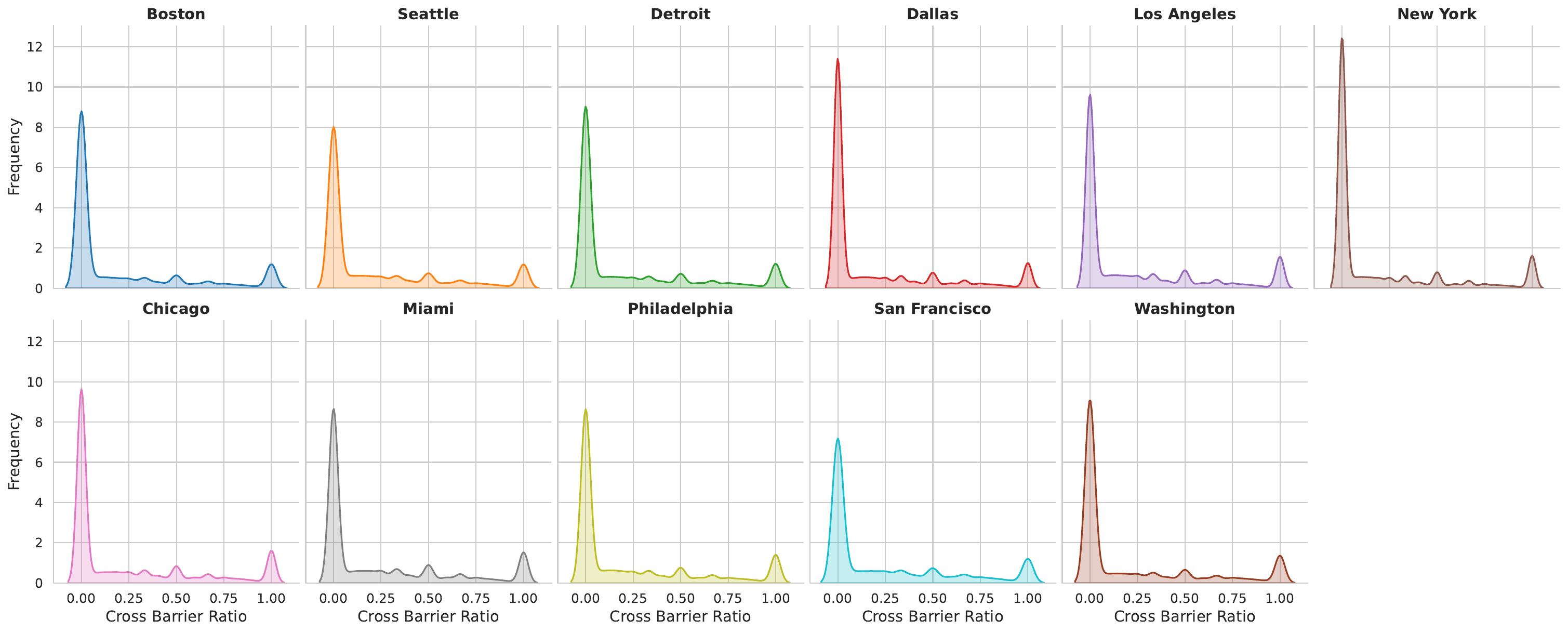}
    \caption[Distribution the user-level cross-barrier ratio (CBR) across the 11 CBSAs analysed]{
    Distribution of the user-level cross-barrier ratio (CBR) across the 11 CBSAs analysed.  
    The pronounced spike at zero indicates that the majority of users do not cross a detected barrier during the observation period.
    }
    \label{supp_fig:CBR_dist}
\end{figure}

Fig.~\ref{fig:4}a summarizes the workflow used to compute the cross-barrier ratio (CBR). Because the origin–destination matrix is highly sparse—only \(3\%\) of CBG pairs register at least one trip in the 2019 baseline—most candidate barriers identified by large positive residuals (top \(5\%\); see Methods) correspond to dyads with zero observed flow.  
To obtain a sample large enough for statistically reliable estimates of the CBR, we therefore  
(i) restrict attention to CBG pairs separated by less than 20 km that exhibit non-zero flow and  
(ii) recompute residuals via $\hat{d}_{ij}^{(e)} \sim \beta \log(d_{ij}^{(p)}) + \epsilon_{ij}$.  
Pairs whose residual now falls in the upper quartile (top \(25\%\)) of this restricted set are labelled soft barriers.  
For every user \(i\) we follow their full trajectories and evaluate the individual CBR, \(r_i\).  
Overall, \(56.5\%\) of users never cross a soft barrier (\(r_i = 0\)) in 2019 study period.  
The CBSA-level distribution of CBR values is displayed in Fig.~\ref{supp_fig:CBR_dist}.

\section{Models}\label{supp:model}

\subsection{Word2Vec}\label{supp:w2v}
We embed mobility trajectories with the skip--gram with negative sampling (SGNS) model implemented in the \texttt{gensim} Python library~\cite{rehurek_lrec}.
All hyper-parameters follow the theoretical correspondence between SGNS and the gravity law of mobility established by Murray \emph{et al.}~\cite{murray2023unsupervised}.
\begin{itemize}
  \item \textbf{Context window.}  
        A symmetric window size of $w=1$ restricts the context to single origin--destination pairs, causing the empirical SGNS objective to coincide with the gravity model’s likelihood~\cite{murray2023unsupervised}.
  \item \textbf{Embedding dimension.}  
        We set the vector dimension to $d = 300$, a value that reliably captures the heterogeneous regularities of large-scale mobility data without undue computational cost~\cite{gu2021principled, murray2023unsupervised}.
  \item \textbf{Minimum frequency.}  
        Spatial tokens occurring fewer than $f_{\min}=50$ times are discarded, removing extremely sparse observations that add noise and slow convergence.
  \item \textbf{Negative samples and Negative-sampling exponent.}  
        The number of noise samples per positive pair is kept at $k = 5$ and the noise distribution exponent is fixed to $\gamma = 0.75$ by default, which balances training speed and representation quality~\cite{mikolov2013efficient}.
\end{itemize}
All remaining parameters follow the \texttt{gensim} defaults.  


\subsection{Logistic Regression Model for Barrier Composition}\label{supp:model_barrier_decompose}
To test the importance of different metrics in creating mobility barriers we used a logistic regression model. Our analysis focuses exclusively on CBG pairs within a 20-kilometer radius. To control for distance-dependent effects, we stratify these pairs into 20 distance bins of 1-kilometer increments. Within each distance group, we construct a balanced binary classification dataset. 
CBG pairs identified as mobility barriers (see Methods) are labeled as the positive class (1), while an equal number of randomly selected CBG pairs from the same distance group serve as the negative class (0). 
This balanced sampling approach ensures robust statistical inference across all distance ranges.

Our regression model incorporates three categories of explanatory variables: POI related variables $\{d_{POI}\}$, physical barrier variables $\{d_{Phy}\}$, and demographic distance variables $\{d_{Demo}\}$. 
Additionally, we include administrative boundary differences $\{d_{ij,\text{County}}\}$ to indicate if the CBG pair is located in different counties. Besides, to obtain the breakdown of the intervening opportunities effect, we train another logistic regression in which the POI intervening opportunities variable (total number of POI between areas) is disaggregated by category while keeping other predictors during the model fitting. 

In our model we use the logarithms of POI intervening opportunity and physical barriers in the model, as they are heavy-tailed (see Fig.~\ref{supp_fig:data_distribution}).
All distance variables are standardized within each distance group to have zero mean and unit variance.  This standardization enables direct comparison of effect magnitudes across variables with different scales and units. 
In logistic regression, standardized coefficients represent the change in log-odds for a one standard deviation increase in the predictor variable. 
For instance, a coefficient of 0.5 for racial distance indicates that a one standard deviation increase in racial dissimilarity increases the odds of barrier formation by a factor of $e^{0.5} = 1.65$. We interpret these standardized coefficients as measures of ``barrier effect.'' 
The sign indicates the direction of association: positive coefficients suggest the variable increases barrier likelihood, while negative coefficients indicate the variable facilitates mobility flow. 
The magnitude reflects the strength of the effect, with larger absolute values indicating more influential determinants of barrier formation.

To evaluate the relative explanatory power of different groups of predictors in our logistic regression model, we use the likelihood ratio test (LRT). The likelihood ratio statistic, $\lambda$, is defined as 
\begin{equation}
    \lambda = -2(\log \mathcal{L}_{\text{full}} - \log \mathcal{L}_{\text{reduced}}),
\end{equation}
represent the log-likelihoods of the full model (which includes all predictors) and the reduced model (which excludes one group of predictors being evaluated), respectively. For instance, in our analysis, comparing the full model to a model that excludes predictors related to physical barriers allows us to quantify the contribution of physical barriers. Similarly, we perform this calculation for predictors related to POIs and demographic variables, enabling us to assess the relative importance of each group of predictors.

\subsection{Linear Regression Model for Cross Barrier Ratio}\label{supp:cross_barrier_ratio}
To examine the relationship between barrier-crossing behavior and residential demographic characteristics, we specify a linear regression model as follows:
\begin{equation}\label{eq:cross_barrier_ratio}
r_{i} = \sum_{j\in \Gamma} \beta^{(j)} X_{i}^{(j)} + \beta^{(c_i)},
\end{equation}
where $r_{i}$ represents the Cross Barrier Ratio for individual $i$, $X_i^{(j)}$ denotes the value of residential demographic variable $j$ for individual $i$, and $\beta^{(j)}$ is the corresponding regression coefficient. 
The model includes a CBSA fixed effect $\beta^{(c_i)}$ to account for metropolitan area-specific characteristics where individual $i$ resides.
The set $\Gamma$ encompasses five key residential demographic variables: employment ratio, public transportation usage ratio for commuting, racial diversity, poverty ratio, and population size. 
These variables capture the socioeconomic and demographic composition of individuals' residential environments, which may influence their propensity to cross mobility barriers.
To address potential residential mobility during the study period, we assigned each individual to their most frequently observed home CBG for demographic characterization. 
This approach ensures that residential demographics reflect the predominant living environment while avoiding overweighting individuals who may have relocated during the observation window in our regression analysis.

\section{Robustness Checks}\label{supp:robustness}
\subsection{Different threshold for trajectory pruning}\label{supp:threshold_robustness}
\begin{figure}[t]
    \centering
    \includegraphics[width=\linewidth]{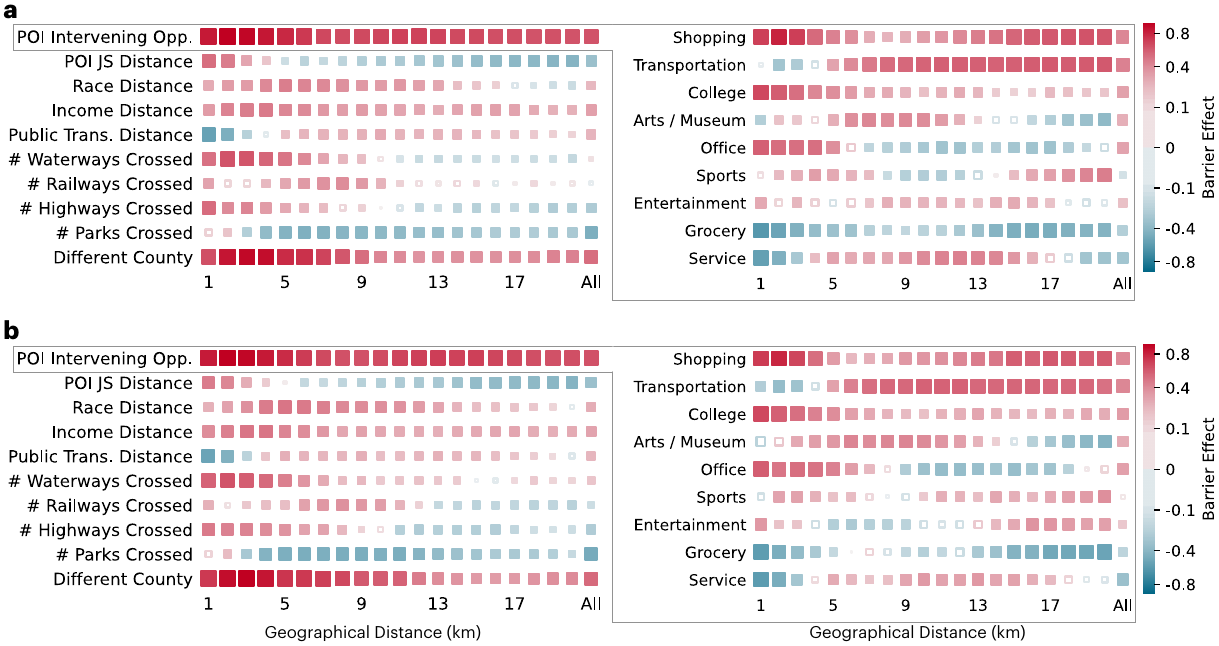}
    \caption[Heatmap of logistic regression coefficients for barrier predictors across distances in H3 (resolution 8) tiles.]{Heatmap of logistic regression coefficients for barrier predictors across distances for trajectory pruning with thresholds \textbf{(a)} $[0.94, 0.995]$  and \textbf{(b)} $[0.94, 0.997]$. 
    Results are presented across 20 distance bins in 11 CBSAs, including POI features, physical infrastructure, demographic differences, and county boundaries.
    Empty symbols correspond to statistically non-significant coefficients ($p > 0.05$).
    The breakdown of the intervening opportunities effect by POI category is shown in the right panel.
    }
    \label{supp_fig:flow_pruning}
\end{figure}

When generating sentences for Word2Vec from trajectories, we constrained flows to lie within the quantile range $[0.95, 0.995]$ to mitigate overfitting issues. 
To assess the sensitivity of our results to alternative threshold choices, we conducted additional analyses constraining flows within the quantile ranges $[0.94, 0.995]$ and $[0.95, 0.997]$. 
The results for the logistic regression analyses are presented in Fig.~\ref{supp_fig:flow_pruning} for both analyses. 
We observe that different thresholds yield very similar results regarding the composition of mobility barriers. 
These findings demonstrate that our main results are robust to the specific choices of trajectory pruning thresholds.

\subsection{Different spacial resolution for identifying mobility barriers}\label{supp:h3_robustness}
The main results in this paper are based on human mobility data at the Census Block Group (CBG) level, which provides the minimum spatial resolution for most demographic data from the U.S. Census while capturing rich human activity patterns. To assess the sensitivity of our results to different spatial resolutions, we conducted additional analyses by aggregating visits to H3 hexagonal tiles at resolution 8 \cite{Uber2018H3}.

Census Block Groups are administrative boundaries created by the U.S. Census Bureau that follow real-world features such as roads, rivers, and political boundaries. CBGs vary significantly in size and shape—city-center tracts can be very small while suburban ones can be very large. They contain between 600 and 3,000 people and serve as the smallest geographic unit for detailed demographic sample data. 
In contrast, H3 tiles are uniform hexagonal cells created by Uber's hierarchical spatial indexing system. H3 divides the Earth into hexagons of consistent size at each resolution level (0-15), with each parent hexagon containing 7 child hexagons at the next resolution level. This creates a predictable, standardized grid covering the entire globe. At resolution 8, the average hexagon area is approximately 0.737 km$^2$.

\begin{figure}[t]
    \centering
    \includegraphics[width=\linewidth]{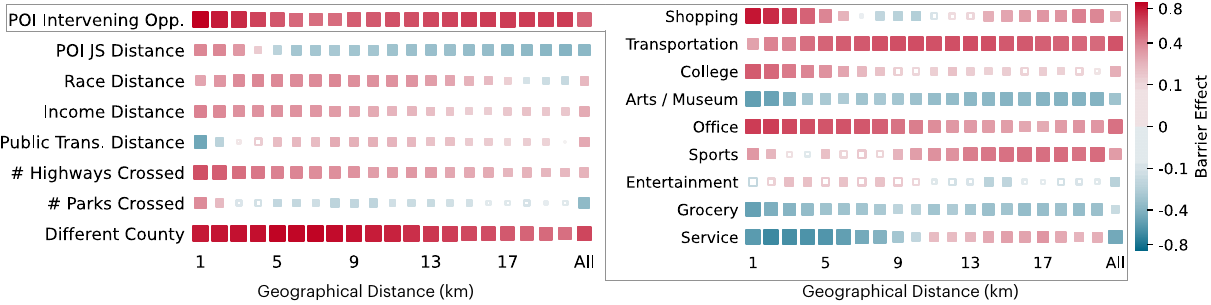}
    \caption[Heatmap of logistic regression coefficients for barrier predictors across distances in H3 (resolution 8) tiles.]{
    Heatmap of logistic regression coefficients for barrier predictors across distances in H3 (resolution 8) tiles.
    Results are presented across 20 distance bins in 11 CBSAs, including POI features, physical infrastructure, demographic differences, and county boundaries.
    Empty symbols correspond to statistically non-significant coefficients ($p > 0.05$).
    The breakdown of the intervening opportunities effect by POI category is shown in the right panel.
    }
    \label{supp_fig:robust_H3}
\end{figure}

To analyze the nature of mobility barriers in H3 tiles, we up-level the visits in H3 (resolution 8) and obtained demographic data by reweighting the data from CBGs based on overlapping area. Other distance measures were obtained following the same approach described in~\ref{supp:distance_measuring}. The results for the logistic regression analyses are presented in Fig.~\ref{supp_fig:robust_H3}. Despite small quantitative differences, we observe that different spatial resolutions yield very similar qualitative results regarding the composition of mobility barriers. These findings demonstrate that our main results are robust to the specific choice of spatial resolution.

\section{Main results by year}\label{supp:barrier_composition_by_year}
\subsection{Understanding the barrier composition}
\begin{figure}[tb]
    \centering
    \includegraphics[width=\linewidth]{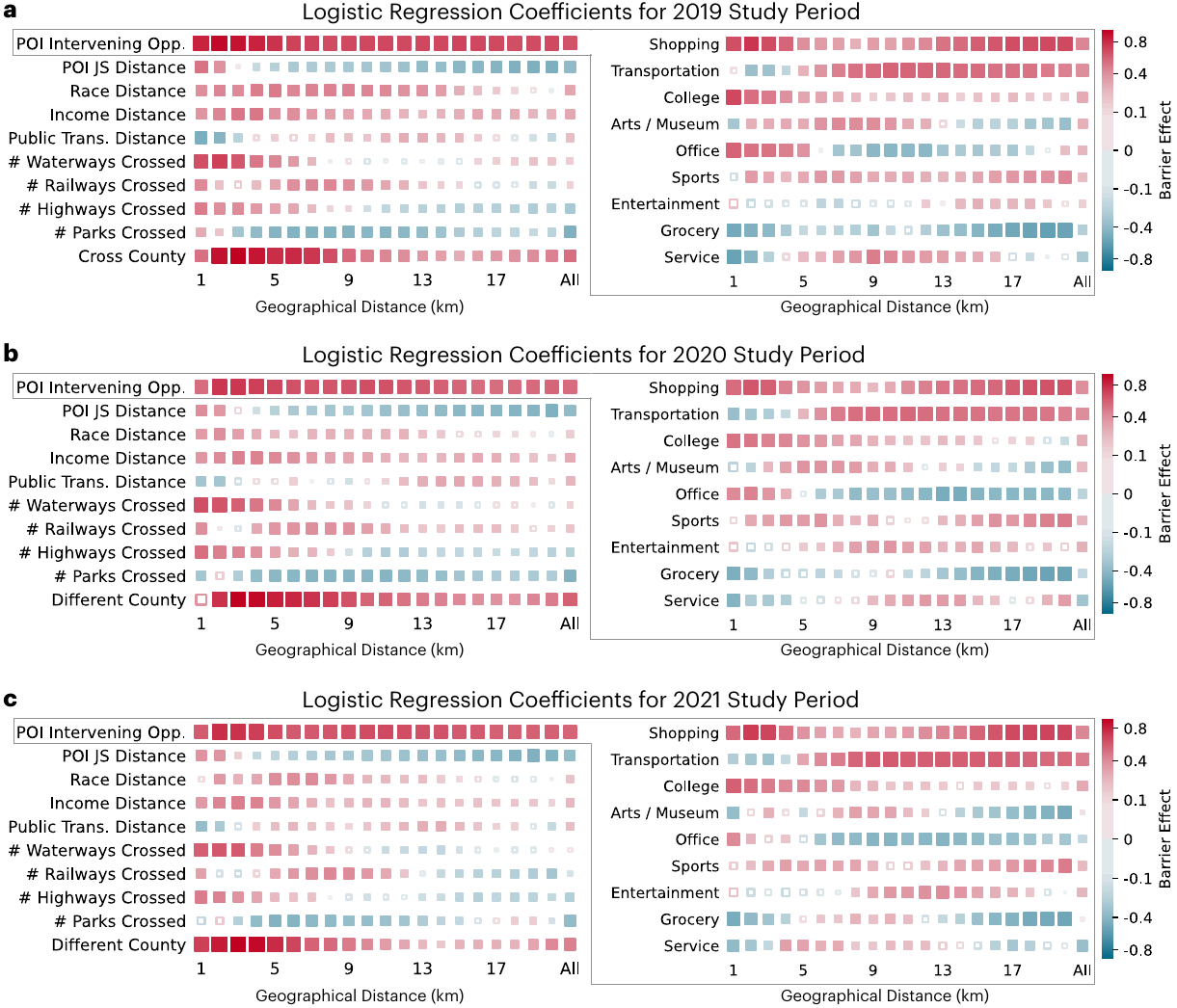}
    \caption[Heatmap of logistic regression coefficients for barrier predictors across distances for years 2019 to 2021.]{
    \textbf{(a), (b) and (c)} 
    Heatmap of logistic regression coefficients for barrier predictors across distances for years 2019 to 2021.
    Results are presented across 20 distance bins in 11 CBSAs, including POI features, physical infrastructure, demographic differences, and county boundaries.
    Empty symbols correspond to statistically non-significant coefficients ($p > 0.05$).
    The breakdown of the intervening opportunities effect by POI category is shown in the right panel.
    }
    \label{supp_fig:main_results_year}
\end{figure}

In this section, we stratify the main results presented in the paper to investigate differences and similarities in the nature of mobility barriers across different years. 
Fig.~\ref{supp_fig:main_results_year} displays the coefficient estimates from Eq.~\eqref{eq:logistic} for the periods 2020-10 to 2021-03 and 2021-10 to 2022-03, aligning with the same temporal framework as the main results from 2019 data presented in the paper. 
Recall that the coefficient estimates represent the change in log-odds of a CBG pair existing a mobility barrier for a one standard deviation change in the predictor variable, which can be interpreted intuitively as the ``barrier effects''.
Our analysis reveals consistencies across the study periods. 
First, we emphasize that most results remain consistent from 2019 to 2021. 
Despite differences in magnitude, the barrier effects maintain the same hierarchical pattern, where POI intervening opportunity remains dominant, followed by demographic differences and physical barriers. 
Additionally, the trends in barrier effects as distance increases remain largely similar across years.
Detailed model fit results for the periods from 2019 to 2021 are presented in Table~\ref{tbl:logistic2019_left},~\ref{tbl:logistic2019_right},~\ref{tbl:logistic2020_left},~\ref{tbl:logistic2020_right}, ~\ref{tbl:logistic2021_left}, and~\ref{tbl:logistic2021_right}.

\subsection{Understanding the cross barrier ratio}\label{supp:cbr_by_year}
The estimation strategy for the cross–barrier ratio is presented in Section~\ref{supp:cross_barrier_ratio}.  
Fig.~\ref{fig:4}d (main text) visualizes the point estimates together with standard errors, while Table~\ref{tbl:cross_barrier_ratio_years} reports the full set of regression coefficients.

\begin{table}[htb]
\centering
\caption[Estimated coefficients of model~\eqref{eq:cross_barrier_ratio} for 2019–2021 study period. ]{
Estimated coefficients of model~\eqref{eq:cross_barrier_ratio} for 2019–2021 study period. 
Robust standard errors are shown in parentheses. 
The reported $p$-values correspond to two–sided tests of the null hypothesis that each coefficient equals zero.
}
\label{tbl:cross_barrier_ratio_years}

\begingroup
\centering
\begin{tabular}{lccc}
   \tabularnewline \midrule \midrule
   Dependent Variable: & \multicolumn{3}{c}{Cross Barrier Ratio}\\
   Year               & 2019            & 2020            & 2021 \\   
   Model:             & (1)             & (2)             & (3)\\  
   \midrule
   \emph{Variables}\\
   Population         & -0.1074$^{***}$ & -0.1178$^{***}$ & -0.1223$^{***}$\\   
                      & (0.0266)        & (0.0273)        & (0.0276)\\   
   Employed Ratio    & 0.3088$^{***}$  & 0.2245$^{**}$   & 0.1866$^{**}$\\   
                      & (0.0661)        & (0.0746)        & (0.0678)\\   
   Poverty Ratio     & -0.1563$^{**}$  & -0.1912$^{**}$  & -0.1373$^{*}$\\   
                      & (0.0653)        & (0.0732)        & (0.0628)\\   
   Racial Diversity  & 0.1380$^{***}$  & 0.1393$^{***}$  & 0.1321$^{***}$\\   
                      & (0.0318)        & (0.0364)        & (0.0416)\\   
   Trans. Public    & 0.2578$^{***}$  & 0.2160$^{***}$  & 0.2924$^{***}$\\   
                      & (0.0492)        & (0.0595)        & (0.0555)\\   
   \midrule
   \emph{Fixed-effects}\\
   CBSA         & Yes             & Yes             & Yes\\  
   \midrule
   \emph{Fit statistics}\\
   Observations       & 8,641,672       & 7,541,528       & 6,782,359\\  
   R$^2$              & 0.03567         & 0.03130         & 0.03377\\  
   Within R$^2$       & 0.03302         & 0.02903         & 0.03154\\  
   \midrule \midrule
   \multicolumn{4}{l}{\emph{Clustered (CBSA) standard-errors in parentheses}}\\
   \multicolumn{4}{l}{\emph{Signif. Codes: ***: 0.01, **: 0.05, *: 0.1}}\\
\end{tabular}
\par\endgroup

\end{table}

\section{Main results by city}\label{supp:cities_normalzied_flux}
\subsection{Modeling normalized flux by city}
We compare the explanatory power of purely geographical distance versus embedding–based distance (cosine distance in the latent space) when modeling the normalized human flux.  
Under the mobility gravity framework $\hat{T}_{ij} = C m_i m_j f(d_{ij})$, the normalized flux between locations \(i\) and \(j\) is
\begin{equation}\label{eq:normalized_flux}
    \frac{\hat{T}_{ij}}{m_i m_j}= C\,f\!\left(d_{ij}\right),
\end{equation}
where \(d_{ij}\) is either the great–circle (geographical) distance or the cosine distance in the embedding space, $f(\cdot)$ is a decay function, \(m_i\) is the number of unique users observed at location \(i\), and \(C\) is a proportionality constant.

\begin{figure}[!tb]
    \centering
    \includegraphics[width=0.8\textwidth]{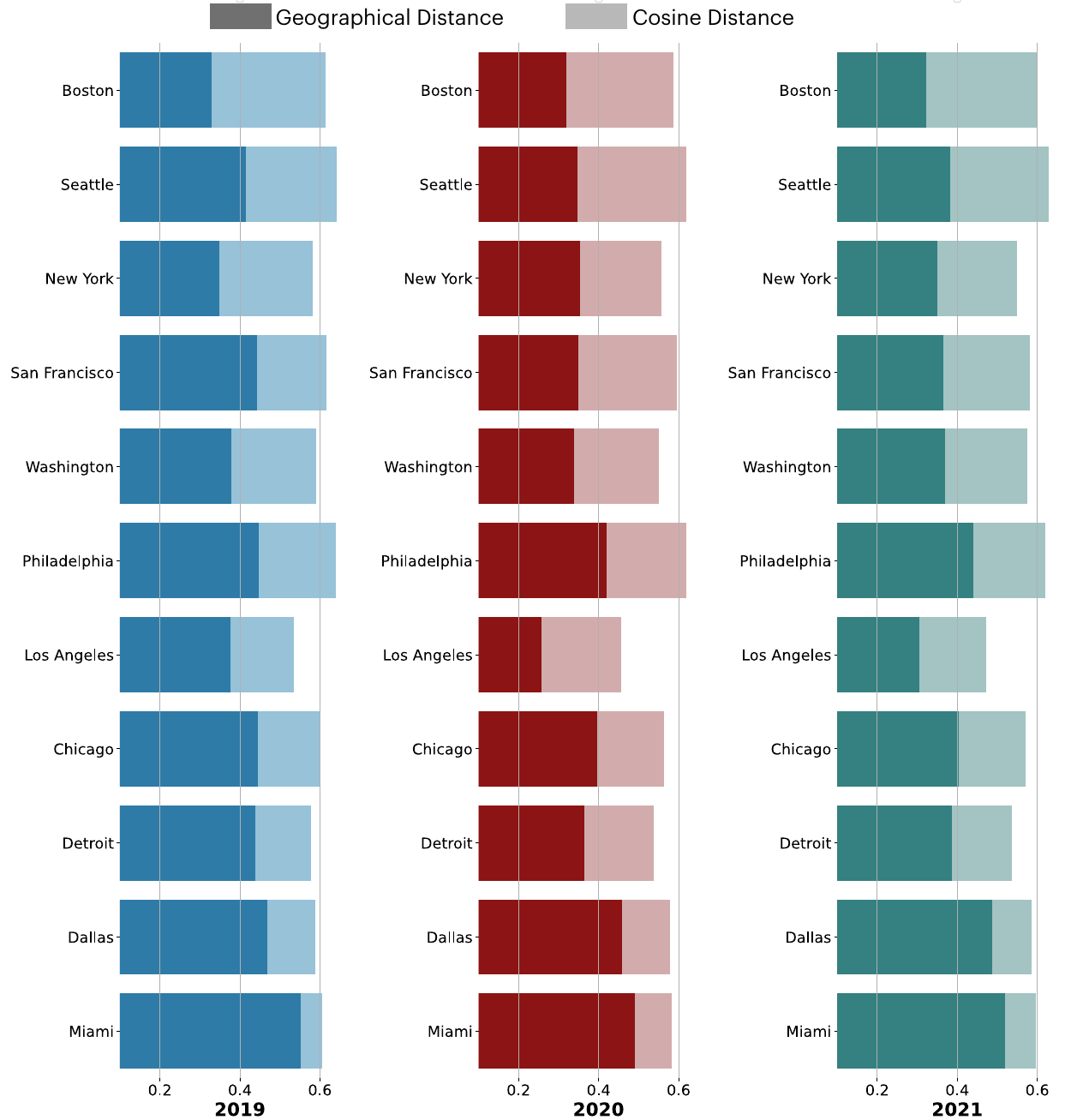}
    \caption[Explanatory power of distance metrics in the normalized gravity model.]{
    Explanatory power of distance metrics in the normalized gravity model.
    For each CBSA–year pair, the darker bar shows the \(R^{2}\) obtained with geographical distance, whereas the lighter bar shows the \(R^{2}\) obtained with embedding cosine distance.  
    Positive gaps indicate that embedding distance yields a better fit.
    }
    \label{supp_fig:cities_normalized_flux}
\end{figure}

Fig.~\ref{supp_fig:cities_normalized_flux} reports the coefficient of determination (\(R^{2}\)) obtained when fitting ~\eqref{eq:normalized_flux} with the two alternative distance metrics across all U.S.\ CBSAs and years.  
Because \(R^{2}\) measures the fraction of the variance in the dependent variable that is accounted for by the model, a higher \(R^{2}\) implies a more accurate explanation of the spatial variation in normalized fluxes.  
On average, substituting geographical distance with embedding cosine distance raises \(R^{2}\) by \(0.185\).  
This means the embedding distance enables the model to explain an additional \(18.5\%\) of the variance in normalized human flux compared with using geographical distance alone, highlighting the superior and temporally robust descriptive power of the learned embeddings across heterogeneous metropolitan areas.

\subsection{Understanding the barrier nature by city}\label{supp:barrier_composition_by_city}
\begin{figure}[tbp]
    \centering
    \includegraphics[width=\linewidth]{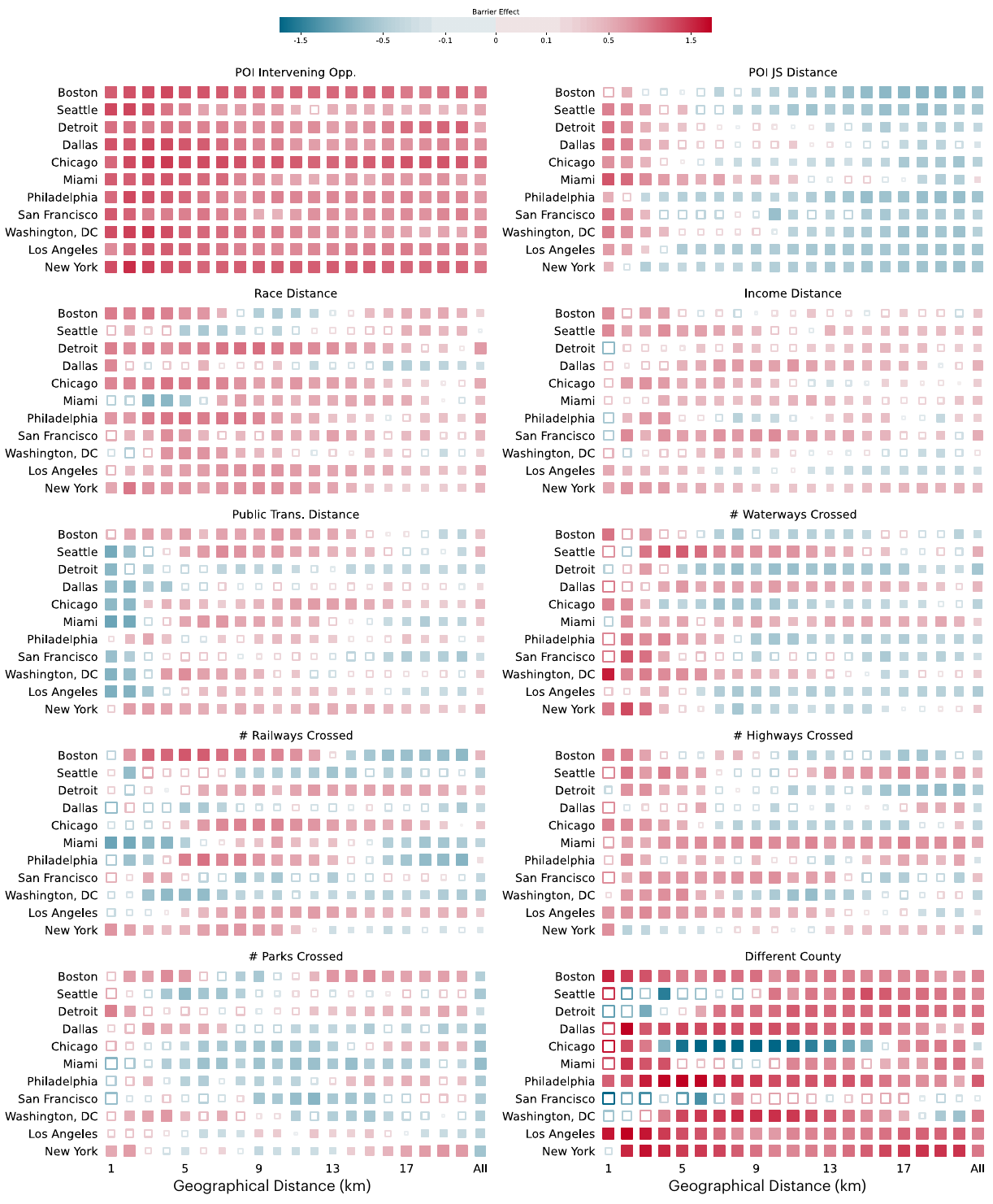}
    \caption[Heatmap of logistic regression coefficients for barrier predictors in each CBSA.]{
    Heatmap of logistic regression coefficients for barrier predictors in each CBSA.
    Results are presented across 20 distance bins for each CBSAs, including POI features, physical infrastructure, demographic differences, and county boundaries.
    Empty symbols correspond to statistically non-significant coefficients ($p > 0.05$).
    The breakdown of the intervening opportunities effect by POI category is shown in the right panel.
    }
    \label{supp_fig:main_results_cities}
\end{figure}

In this section, we examine the differences and similarities in the nature of mobility barriers across the 11 CBSAs included in our study. Fig.~\ref{supp_fig:main_results_cities} presents the logistic regression coefficients across 11 U.S. CBSAs, showing how different factors contribute to mobility barriers over 20 distance bins. Overall, we find that the key patterns identified in the main paper hold consistently across urban areas, reinforcing the generality and robustness of our results. In particular:
\begin{itemize}
    \item Amenity structure emerges as the primary determinant of mobility barriers across all areas. POI intervening opportunities exhibit large positive associations with barrier formation in all cities and all distances, while across cities, POI Jensen-Shannon (JS) distance shows positive association with barriers at short distances and negative association at large distances.
    \item Administrative boundaries constitute the second strongest predictor of mobility barriers across all metropolitan areas. Although the effect is largely positive, there are some small deviations in cities like Chicago and San Francisco, where the effect of administrative boundaries is more muted or distance-dependent.
    \item Demographic differences exhibit significant positive associations with barrier presence across all areas and most distances. However, we can see some differences in cities like Miami, where Race distance is positive at small distances.
    \item Physical barriers such as highways, railways, and parks also display common patterns across cities. Highways consistently act as barriers at short distances but tend to weaken or reverse at longer ranges, reflecting their dual role as both local obstacles and regional connectors. Parks are positively associated with barriers in cities like Boston and New York, particularly at short distances, but have neutral or even negative effects in Miami, suggesting that their function as barriers or facilitators depends on local context. Railways tend to act as barriers in most cities, except in Washington, Seattle, or Boston, where they show a negative association with barriers, likely due to the city's integrated rail system.
\end{itemize}

Thus, despite the results in the main paper are prevalent across cities, these metropolitan-specific variations underscore the importance of local context in shaping the relationship between urban infrastructure and mobility patterns, highlighting the need for place-based approaches to understanding and addressing urban mobility barriers.

\subsection{Understanding the cross barrier ratio}
\begin{figure}[t]
    \centering
    \includegraphics[width=\linewidth]{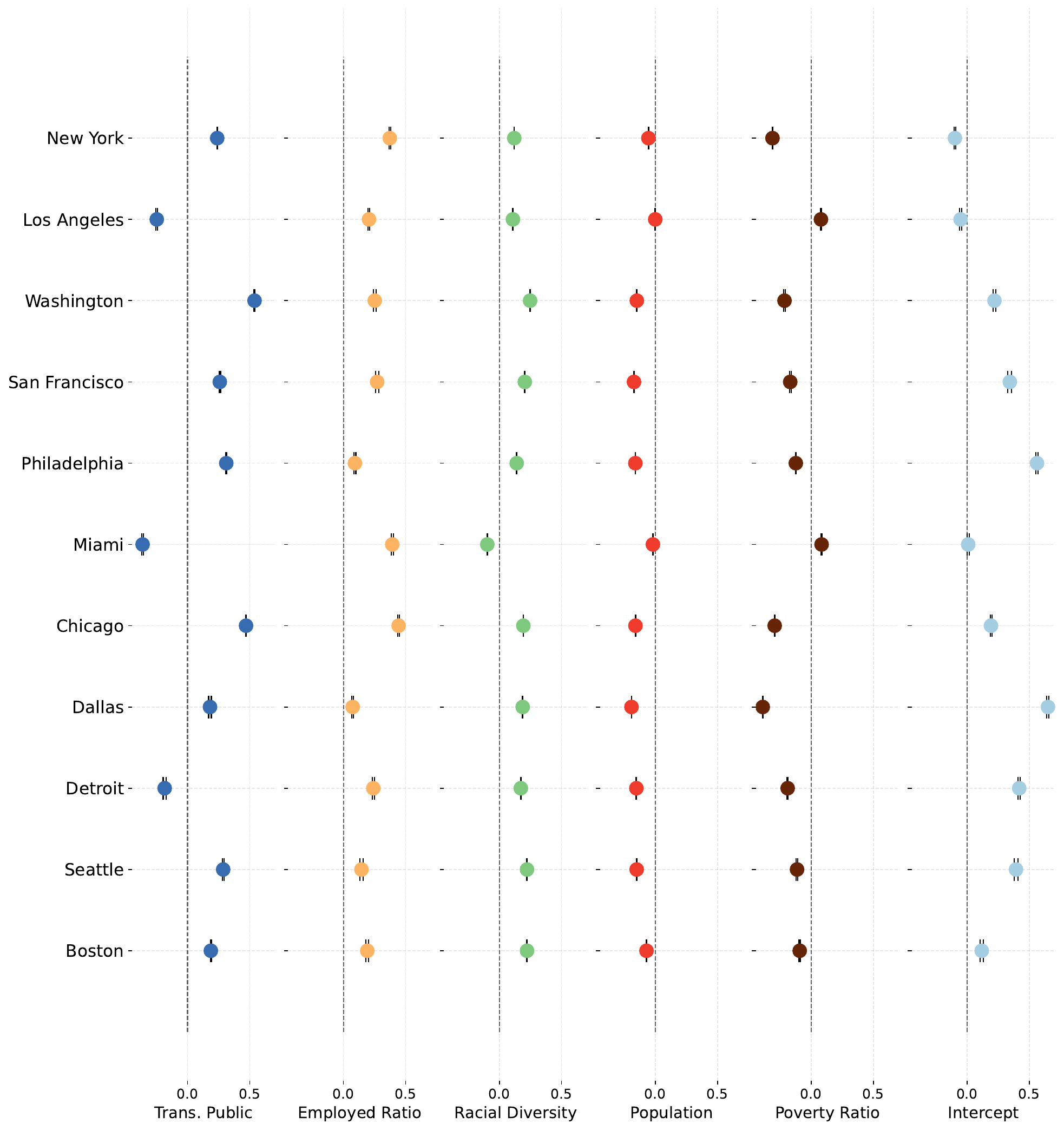}
    \caption[Estimated coefficients of model~\eqref{eq:cross_barrier_ratio} for different CBSA in 2019 study period.]{
    Estimated coefficients of model~\eqref{eq:cross_barrier_ratio} for different CBSA in 2019 study period. 
    Error bars denote standard errors of coefficient estimates.
    }
    \label{supp_fig:cross_barrier_coeff_2019}
\end{figure}

The estimation protocol for the cross–barrier ratio is detailed in Section~\ref{supp:cross_barrier_ratio}. Fig.~\ref{supp_fig:cross_barrier_coeff_2019} depicts the 2019 point estimates with standard errors by city, whereas Table~\ref{tbl:cross_barrier_ratio_cities} lists the complete set of regression coefficients by city. As we can see, our results in the main paper are reproduced qualitatively by city. In particular, individuals living in areas with higher Employment ration, more Public-transport usage, and more Racial diversity tend to cross barriers with more frequency. Notable exceptions are Miami and Los Angeles, where Public Transportation usage is related to less barrier-crossing, possibly reflecting different transit usage patterns or network coverage in those cities. Conversely, residents of areas with higher poverty rates or larger populations are generally less likely to cross barriers, consistent with reduced mobility access. While the magnitude of effects varies across cities, the direction and structure of associations remain consistent, reinforcing the robustness of our findings and the behavioral relevance of demographic context in shaping cross-barrier mobility.

\section{Libraries used}\label{supp:library}
Analysis was conducted in Python and R using the following packages:
\begin{enumerate}
    \item Python Package \texttt{gensim}~\cite{rehurek_lrec} v4.3.0 for producing CBG embeddings
    \item Python Package \texttt{Umap}~\cite{McInnes2018} for dimension reduction for CBG embeddings.
    \item Python Package \texttt{Pandas}~\cite{mckinney-proc-scipy-2010} for loading, transforming, and analyzing data tables.
    \item Python Package \texttt{GeoPandas}~\cite{kelsey_jordahl_2020_3946761} for spatial analysis and plotting map figures.
    \item Python Package \texttt{statsmodels}~\cite{seabold2010statsmodels} and {\tt SciPy}~\cite{Virtanen2020SciPy} for statistical modeling analysis.
    \item Python Package \texttt{OSMnx}~\cite{boeing2025modeling} v1.9.4 for accessing physical barrier polygons from Open Street Map.
    \item Python Package {\tt matplotlib}~\cite{Hunter:2007}, \texttt{seaborn}~\cite{Waskom2021}, and R package \texttt{ggplot2}~\cite{ggplot2}  were used for the visualizations.
    \item R package {\tt fixest}~\cite{fixest} for logistic and linear regressions.
    \item R package \texttt{tidycensus}~\cite{tidycensus} for accessing the Census data. 
    \item R package \texttt{tigris}~\cite{tigris} for accessing boundaries of the Census Block Groups and counties.
\end{enumerate}


\clearpage

\begin{sidewaystable}[htb]                    
  \centering
  \caption[Estimated coefficients of model given by Eq.~\eqref{eq:logistic} for the 2019 study period.]{
    Estimated coefficients of model given by Eq.~\eqref{eq:logistic} for the 2019 study period with respective standard errors in parentheses. 
    P-values correspond to two-sided tests for the hypothesis that each coefficient differs from zero. 
    We also report the Squared Correlation and Pseudo $R^2$ results for the logistic regression, and the training accuracy with its 95\% confidence interval in square brackets.
    }
  \label{tbl:logistic2019_left}
  \adjustbox{max width=\textheight,center}{   
\begin{tabular}{lccccccccccccccccccccc}
   \tabularnewline \midrule \midrule
   Dependent Variable: & \multicolumn{21}{c}{label}\\
   Distance (km)          & 1                               & 2                               & 3                               & 4                               & 5                               & 6                               & 7                               & 8                               & 9                               & 10                           & 11                              & 12                              & 13                              & 14                              & 15                              & 16                              & 17                              & 18                              & 19                              & 20                              & All \\   
   Model:                 & (1)                             & (2)                             & (3)                             & (4)                             & (5)                             & (6)                             & (7)                             & (8)                             & (9)                             & (10)                         & (11)                            & (12)                            & (13)                            & (14)                            & (15)                            & (16)                            & (17)                            & (18)                            & (19)                            & (20)                            & (21)\\  
   \midrule
   \emph{Variables}\\
\end{tabular}
  }
\end{sidewaystable}

\begin{sidewaystable}[htb]                    
  \centering
  \caption[Estimated coefficients of model given by Eq.~\eqref{eq:logistic} for the 2019 study period with disaggregated POI intervening opportunity.]{
    Estimated coefficients of model given by Eq.~\eqref{eq:logistic} for the 2019 study period with disaggregated POI intervening opportunity. 
    The respective standard errors are in parentheses
    P-values correspond to two-sided tests for the hypothesis that each coefficient differs from zero. 
    We also report the Squared Correlation and Pseudo $R^2$ results for the logistic regression, and the training accuracy with its 95\% confidence interval in square brackets.
    }
    \label{tbl:logistic2019_right}
  \adjustbox{max width=\textheight,center}{   
\begin{tabular}{lccccccccccccccccccccc}
   \tabularnewline \midrule \midrule
   Dependent Variable: & \multicolumn{21}{c}{label}\\
   Distance (km)          & 1                               & 2                               & 3                               & 4                               & 5                               & 6                              & 7                               & 8                               & 9                               & 10                              & 11                              & 12                             & 13                              & 14                             & 15                              & 16                              & 17                              & 18                              & 19                              & 20                              & All \\   
   Model:                 & (1)                             & (2)                             & (3)                             & (4)                             & (5)                             & (6)                            & (7)                             & (8)                             & (9)                             & (10)                            & (11)                            & (12)                           & (13)                            & (14)                           & (15)                            & (16)                            & (17)                            & (18)                            & (19)                            & (20)                            & (21)\\  
   \midrule
   \emph{Variables}\\
\end{tabular}
  }
\end{sidewaystable}

\begin{sidewaystable}[htb]                    
  \centering
    \caption[Estimated coefficients of model given by Eq.~\eqref{eq:logistic} for the 2020 study period.]{
    Estimated coefficients of model given by Eq.~\eqref{eq:logistic} for the 2020 study period with respective standard errors in parentheses. 
    P-values correspond to two-sided tests for the hypothesis that each coefficient differs from zero. 
    We also report the Squared Correlation and Pseudo $R^2$ results for the logistic regression, and the training accuracy with its 95\% confidence interval in square brackets.
    }
    \label{tbl:logistic2020_left}
  \adjustbox{max width=\textheight,center}{   
\begin{tabular}{lccccccccccccccccccccc}
   \tabularnewline \midrule \midrule
   Dependent Variable: & \multicolumn{21}{c}{label}\\
   Distance (km)          & 1                               & 2                               & 3                               & 4                               & 5                               & 6                               & 7                               & 8                               & 9                               & 10                              & 11                              & 12                              & 13                              & 14                              & 15                              & 16                              & 17                            & 18                            & 19                             & 20                              & All \\   
   Model:                 & (1)                             & (2)                             & (3)                             & (4)                             & (5)                             & (6)                             & (7)                             & (8)                             & (9)                             & (10)                            & (11)                            & (12)                            & (13)                            & (14)                            & (15)                            & (16)                            & (17)                          & (18)                          & (19)                           & (20)                            & (21)\\  
   \midrule
   \emph{Variables}\\
\end{tabular}
  }
\end{sidewaystable}

\begin{sidewaystable}[htb]                    
  \centering
    \caption[Estimated coefficients of model given by Eq.~\eqref{eq:logistic} for the 2020 study period with disaggregated POI intervening opportunity.]{
    Estimated coefficients of model given by Eq.~\eqref{eq:logistic} for the 2020 study period  with disaggregated POI intervening opportunity.
    The respective standard errors are in parentheses. 
    P-values correspond to two-sided tests for the hypothesis that each coefficient differs from zero. 
    We also report the Squared Correlation and Pseudo $R^2$ results for the logistic regression, and the training accuracy with its 95\% confidence interval in square brackets.
    }
    \label{tbl:logistic2020_right}
  \adjustbox{max width=\textheight,center}{   
\begin{tabular}{lccccccccccccccccccccc}
   \tabularnewline \midrule \midrule
   Dependent Variable: & \multicolumn{21}{c}{label}\\
   Distance (km)          & 1                               & 2                              & 3                               & 4                               & 5                               & 6                               & 7                               & 8                               & 9                              & 10                              & 11                              & 12                              & 13                              & 14                              & 15                              & 16                              & 17                             & 18                              & 19                              & 20                              & All \\   
   Model:                 & (1)                             & (2)                            & (3)                             & (4)                             & (5)                             & (6)                             & (7)                             & (8)                             & (9)                            & (10)                            & (11)                            & (12)                            & (13)                            & (14)                            & (15)                            & (16)                            & (17)                           & (18)                            & (19)                            & (20)                            & (21)\\  
   \midrule
   \emph{Variables}\\
\end{tabular}
  }
\end{sidewaystable}

\begin{sidewaystable}[htb]                    
  \centering
    \caption[Estimated coefficients of model given by Eq.~\eqref{eq:logistic} for the 2021 study period.]{
    Estimated coefficients of model given by Eq.~\eqref{eq:logistic} for the 2021 study period with respective standard errors in parentheses. 
    P-values correspond to two-sided tests for the hypothesis that each coefficient differs from zero. 
    We also report the Squared Correlation and Pseudo $R^2$ results for the logistic regression, and the training accuracy with its 95\% confidence interval in square brackets.
    }
    \label{tbl:logistic2021_left}
  \adjustbox{max width=\textheight,center}{   
\begin{tabular}{lccccccccccccccccccccc}
   \tabularnewline \midrule \midrule
   Dependent Variable: & \multicolumn{21}{c}{label}\\
   Distance (km)          & 1                               & 2                               & 3                               & 4                               & 5                               & 6                               & 7                               & 8                               & 9                               & 10                              & 11                              & 12                              & 13                             & 14                              & 15                              & 16                              & 17                             & 18                              & 19                              & 20                              & All \\   
   Model:                 & (1)                             & (2)                             & (3)                             & (4)                             & (5)                             & (6)                             & (7)                             & (8)                             & (9)                             & (10)                            & (11)                            & (12)                            & (13)                           & (14)                            & (15)                            & (16)                            & (17)                           & (18)                            & (19)                            & (20)                            & (21)\\  
   \midrule
   \emph{Variables}\\
\end{tabular}
  }
\end{sidewaystable}

\begin{sidewaystable}[htb]                    
  \centering
    \caption[Estimated coefficients of model given by Eq.~\eqref{eq:logistic} for the 2021 study period with disaggregated POI intervening opportunity.]{
    Estimated coefficients of model given by Eq.~\eqref{eq:logistic} for the 2021 study period  with disaggregated POI intervening opportunity.
    The respective standard errors are in parentheses. 
    P-values correspond to two-sided tests for the hypothesis that each coefficient differs from zero. 
    We also report the Squared Correlation and Pseudo $R^2$ results for the logistic regression, and the training accuracy with its 95\% confidence interval in square brackets.
    }
    \label{tbl:logistic2021_right}
  \adjustbox{max width=\textheight,center}{   
\begin{tabular}{lccccccccccccccccccccc}
   \tabularnewline \midrule \midrule
   Dependent Variable: & \multicolumn{21}{c}{label}\\
   Distance (km)          & 1                               & 2                               & 3                               & 4                              & 5                               & 6                              & 7                               & 8                               & 9                               & 10                           & 11                             & 12                              & 13                              & 14                              & 15                              & 16                              & 17                             & 18                              & 19                              & 20                             & All \\   
   Model:                 & (1)                             & (2)                             & (3)                             & (4)                            & (5)                             & (6)                            & (7)                             & (8)                             & (9)                             & (10)                         & (11)                           & (12)                            & (13)                            & (14)                            & (15)                            & (16)                            & (17)                           & (18)                            & (19)                            & (20)                           & (21)\\  
   \midrule
   \emph{Variables}\\
\end{tabular}
  }
\end{sidewaystable}

\begin{sidewaystable}[tb]                    
  \centering
    \caption[Estimated coefficients of model~\eqref{eq:cross_barrier_ratio} across 11 CBSA in 2019 study period.]{
        Estimated coefficients of model~\eqref{eq:cross_barrier_ratio} across 11 CBSA in 2019 study period. 
        Robust standard errors are shown in parentheses. 
        The reported $p$-values correspond to two–sided tests of the null hypothesis that each coefficient equals zero.
    }
    \label{tbl:cross_barrier_ratio_cities}
  \adjustbox{max width=\textheight,center}{   
     
\begingroup
\centering
\begin{tabular}{lccccccccccc}
   \tabularnewline \midrule \midrule
   Dependent Variable: & \multicolumn{11}{c}{Cross Barrier Ratio}\\
   CBSA               & Boston          & Seattle         & Philadelphia    & San Francisco   & Washington      & Detroit         & Dallas          & Miami           & Chicago         & Los Angeles     & New York \\   
   Model:             & (1)             & (2)             & (3)             & (4)             & (5)             & (6)             & (7)             & (8)             & (9)             & (10)            & (11)\\  
   \midrule
   \emph{Variables}\\
   Constant           & 0.1173$^{***}$  & 0.3932$^{***}$  & 0.5620$^{***}$  & 0.3434$^{***}$  & 0.2198$^{***}$  & 0.4190$^{***}$  & 0.6499$^{***}$  & 0.0087          & 0.1920$^{***}$  & -0.0524$^{***}$ & -0.0982$^{***}$\\   
                      & (0.0135)        & (0.0150)        & (0.0090)        & (0.0155)        & (0.0125)        & (0.0093)        & (0.0082)        & (0.0088)        & (0.0063)        & (0.0079)        & (0.0074)\\   
   POPULATION         & -0.0708$^{***}$ & -0.1497$^{***}$ & -0.1602$^{***}$ & -0.1713$^{***}$ & -0.1483$^{***}$ & -0.1519$^{***}$ & -0.1913$^{***}$ & -0.0193$^{***}$ & -0.1593$^{***}$ & -0.0002         & -0.0551$^{***}$\\   
                      & (0.0024)        & (0.0026)        & (0.0018)        & (0.0025)        & (0.0019)        & (0.0018)        & (0.0011)        & (0.0014)        & (0.0012)        & (0.0013)        & (0.0014)\\   
   EMPLOYED\_RATIO    & 0.1916$^{***}$  & 0.1450$^{***}$  & 0.0922$^{***}$  & 0.2711$^{***}$  & 0.2518$^{***}$  & 0.2404$^{***}$  & 0.0745$^{***}$  & 0.3926$^{***}$  & 0.4440$^{***}$  & 0.2056$^{***}$  & 0.3727$^{***}$\\   
                      & (0.0116)        & (0.0129)        & (0.0075)        & (0.0137)        & (0.0110)        & (0.0079)        & (0.0076)        & (0.0078)        & (0.0054)        & (0.0070)        & (0.0061)\\   
   POVERTY\_RATIO     & -0.0923$^{***}$ & -0.1138$^{***}$ & -0.1237$^{***}$ & -0.1686$^{***}$ & -0.2146$^{***}$ & -0.1893$^{***}$ & -0.3898$^{***}$ & 0.0835$^{***}$  & -0.2937$^{***}$ & 0.0784$^{***}$  & -0.3112$^{***}$\\   
                      & (0.0053)        & (0.0061)        & (0.0038)        & (0.0066)        & (0.0056)        & (0.0039)        & (0.0028)        & (0.0037)        & (0.0030)        & (0.0031)        & (0.0028)\\   
   RACIAL\_DIVERSITY  & 0.2225$^{***}$  & 0.2222$^{***}$  & 0.1397$^{***}$  & 0.2048$^{***}$  & 0.2473$^{***}$  & 0.1724$^{***}$  & 0.1875$^{***}$  & -0.0971$^{***}$ & 0.1931$^{***}$  & 0.1091$^{***}$  & 0.1198$^{***}$\\   
                      & (0.0016)        & (0.0017)        & (0.0012)        & (0.0023)        & (0.0014)        & (0.0014)        & (0.0010)        & (0.0011)        & (0.0009)        & (0.0010)        & (0.0010)\\   
   PUBLIC\_COMMUTE    & 0.1875$^{***}$  & 0.2867$^{***}$  & 0.3116$^{***}$  & 0.2594$^{***}$  & 0.5389$^{***}$  & -0.1838$^{***}$ & 0.1807$^{***}$  & -0.3610$^{***}$ & 0.4706$^{***}$  & -0.2472$^{***}$ & 0.2388$^{***}$\\   
                      & (0.0041)        & (0.0061)        & (0.0038)        & (0.0062)        & (0.0040)        & (0.0118)        & (0.0109)        & (0.0080)        & (0.0025)        & (0.0054)        & (0.0013)\\   
   \midrule
   \emph{Fit statistics}\\
   Observations       & 411,383         & 423,209         & 684,177         & 345,041         & 599,407         & 575,355         & 1,156,189       & 827,512         & 1,175,402       & 1,210,867       & 1,233,130\\  
   R$^2$              & 0.07641         & 0.05298         & 0.04052         & 0.04029         & 0.08519         & 0.03911         & 0.05625         & 0.01700         & 0.08107         & 0.01287         & 0.04874\\  
   Adjusted R$^2$     & 0.07640         & 0.05297         & 0.04052         & 0.04028         & 0.08518         & 0.03910         & 0.05625         & 0.01699         & 0.08107         & 0.01287         & 0.04874\\  
   \midrule \midrule
   \multicolumn{12}{l}{\emph{IID standard-errors in parentheses}}\\
   \multicolumn{12}{l}{\emph{Signif. Codes: ***: 0.01, **: 0.05, *: 0.1}}\\
\end{tabular}
\par\endgroup

  }
\end{sidewaystable}

\end{document}